\documentclass[showpacs,twocolumn]{svjour3}
\usepackage{amsmath, amssymb}
\usepackage{graphicx}
\usepackage{dcolumn}
\usepackage{subfigure}

\journalname{Brazilian Journal of Physics}

\begin{document}

\title{\sc Geometry in the entanglement dynamics of the double Jaynes--Cummings model.}

\author{A. R. Vieira         \and
        J. G. G. de Oliveira Junior \and
J. G. Peixoto de Faria \and M. C. Nemes }

\institute{A. R. Vieira \at Departamento de F\'{\i}sica - CP 702 - Universidade Federal de Minas Gerais - 30123-970 - Belo Horizonte - MG - Brazil\\
\email{arvieira@fisica.ufmg.br} \and J. G. G. de Oliveira Junior \at Centro de Forma\c c\~ao de Professores, Universidade Federal do
Rec\^oncavo da Bahia, 45.300-000, Amargosa, BA, Brazil\\ \email{zgeraldo@ufrb.edu.br} \and J. G. Peixoto de Faria \at Departamento Acad\^emico de Disciplinas B\'
asicas - Centro Federal de Educa\c c\~ao Tecnol\'ogica de Minas Gerais - 30510-000 - Belo Horizonte - MG - Brazil\\ \email{jgpfaria@des.cefetmg.br} \and M. C. Nemes 
\at Departamento de F\'{\i}sica - CP 702 - Universidade Federal de Minas Gerais - 30123-970 - Belo Horizonte - MG - Brazil\\ \email{carolina@fisica.ufmg.br}}

\date{Received: date / Accepted: date}

\maketitle

\begin{abstract}
We report on the geometric character of the entanglement dynamics of
two pairs of {\it {\it qubits}} evolving according to the double Jaynes--Cummings
model. We show that the entanglement dynamics for the initial states
$|\psi_0\rangle=\cos\alpha|10\rangle+\sin\alpha|01\rangle$ and
$|\phi_0\rangle=\cos\alpha|11\rangle+\sin\alpha|00\rangle$ cover
3--dimensional surfaces in the diagram $C_{ij} \,\mathrm{x}\, C_{ik}
\,\mathrm{x}\, C_{il}$, where $C_{mn}$ stands for the concurrence
between {\it qubits} $m$ and $n$, varying $0\leq\alpha\leq\pi/2$. In the
first case projections of the surfaces on a diagram $C_{ij}
\,\mathrm{x}\, C_{kl}$ are conics. In the second case curves can
be more complex. We relate those conics with a measurable quantity, the
{\it predictability}. We also derive inequalities limiting the sum of the
squares of the concurrence of every bipartition and show that sudden
death of entanglement is intimately connected to the size of the
average radius of a hyper-sphere.
\end{abstract}

\section{Introduction}

The capacity of quantum systems to entangle is perhaps the most
intriguing aspect of quantum mechanics and is a feature that
distinguishes classical from quantum physics. In a seminal
work, Einstein, Podolsky, and Rosen \cite{epr} have brought this
property to discussion and since then the subject has been 
investigated. Recently, pure bipartite interacting quantum systems
have proven to be a very useful tool to explore entanglement dynamics
and unveil several of the intriguing properties which govern quantum
correlations exchange. Examples of such properties are the
{ sudden (or asymptotic) disappearance of entanglement} \cite{morte}, the
so called { entanglement sudden birth} \cite{nascimento}, { control of
entanglement dynamics} \cite{controle} and { entanglement distribution}
\cite{distribuicao1}, an important ingredient for quantum computation.
Perhaps the best known and explored model is the 
Jaynes--Cummings Model (JCM) \cite{JCM}, where several dynamical
scenarios have been explored both with and without dissipation. An
analogous model, the Tavis-Cummings model \cite{tc} has also been
used for similar purposes. The result
obtained in these two contexts have enlightened entanglement
disappearance in finite time \cite{morte1,morte2,morte3}, {relations
between purity, energy and entanglement} \cite{energia,energia1},
invariant entanglement \cite{distribuicao} and { general aspects of
entanglement dynamics between partitions}
\cite{geral,geral1,geral2,geral3}. In the present work we show that
the entanglement dynamics of the Double Jaynes--Cummings Model
(DJCM) \cite{morte1} exhibits geometric properties for
the two classes of initial states we considered. The scenario
is a pair of initially entangled non-interacting atoms $``A"$ and
$``B"$, two cavities ``$a$" and ``$b$" which interact
locally via the JCM and we use concurrence \cite{wootters2} to quantify
entanglement between these parts. We show that, for initial atomic states belonging to
the class
$|\psi_0\rangle=\cos\alpha|10\rangle+\sin\alpha|01\rangle$, the
relations between concurrences describe a conic in a diagram $C_{ij}
\,\mathrm{x}\, C_{kl}$, with $ij \neq kl$ ( $ij$ being equal to
$Aa$, $Ab$, $AB$, $ab$, $aB$ and $Bb$). On the other hand, if the
initial atomic state belongs to the class
$|\phi_0\rangle=\cos\alpha|11\rangle+\sin\alpha|00\rangle$, the
geometric curve is not as simple. However, in all cases when a conic
is found, the eccentricity can be written as a function of the
absolute value of the average excitations in $A$, in other words:
$\mathcal{P}_0=\Bigl|\mathrm{tr}\bigl(\sigma_{z}^{A}\rho_0\bigr)\Bigr|$.
If the initial atomic state is $|\psi_0\rangle$, $\mathcal{P}_0$
gives the probability of the excitation being found in only one of
the two bipartition $Aa$ or $Bb$. On the other hand, if the initial
state is $|\phi_0\rangle$, $\mathcal{P}_0$ does not have the same
interpretation. It is important to notice that $\mathcal{P}_0$ is
the {\it predictability} which according to the complementarity
relation between two {\it qubits} proposed in ref. \cite{bergou} is
related to the initial concurrence. We find that this geometric character
can be extended for more dimensions. It is possible to define a
hyper-surface over which the concurrence dynamics between
every two pairs $i$ and $j$ defines a trajectory over or inside this
hyper surface.

The present work is organized as follows: In section \ref{mf} we
present the physical model and the time evolution for the two classes
of states, $|\psi_0\rangle$ and $|\phi_0\rangle$ ; Next, in section
\ref{Din_emara}, we determine the entanglement (quantified by
concurrence), and we construct the diagram $C_{ij} \,\mathrm{x}\,
C_{kl}$ showing that whenever a conic is found its eccentricity is
related to the {\it predictability} as defined in \cite{bergou}; In
the following section, we show the existence of an
entanglement surface for the dynamics of the pairs of concurrences
involving the same {\it qubit} and justify why curves of the
diagrams $C_{ij} \,\mathrm{x}\, C_{kl}$ will be over that surface;
In section \ref{hiper_esfera} we find an inequality which describes
the entanglement dynamics of all qubit pairs; In section \ref{sec6}
we present how decoherence affects some of the conics and we conclude
in section \ref{conclusion}.

\section{The physical model}\label{mf}

Consider a composite system of two identical two-level atoms (``$A$"
e ``$B$") and two identical cavities (``$a$" e ``$b$"). The atom
``$A$" (``$B$") interacts resonantly with the cavity ``$a$"
(``$b$"), respectively, via JCM \cite{JCM} and the evolution of the
system is governed by the Hamiltonian
\begin{eqnarray}
\nonumber
  H &=& \hbar\omega a^\dag a+\hbar\omega b^\dag b +
  \dfrac{\hbar\omega}{2}\sigma_{z}^{A}+\dfrac{\hbar\omega}{2}\sigma_{z}^{B}+\\ && +
g\bigl(a^\dag \sigma_{-}^{A}+a\sigma_{+}^{A}\bigr)+g\bigl(b^\dag
\sigma_{-}^{B}+b\sigma_{+}^{B}\bigr)\label{H},
\end{eqnarray}
where $a^\dag$ ($b^\dag$) and $a$ ($b$) are the {\it creation} and
{\it annihilation} operators of the field inside cavity $a$ ($b$),
respectively. The matrices $\sigma_{-}^{i}$, $\sigma_{+}^{i}$ and
$\sigma_{z}^{i}$ are Pauli matrices of the i--{\it th} atom, with $i=A,B$.
The cavities are resonant with the atoms, i. e. the frequency of the
field inside each cavity is equal to the frequency of the atomic
transition of the atoms' internal levels.

\begin{figure}[!htb]
 \centering
  \includegraphics[scale=0.550,angle=00]{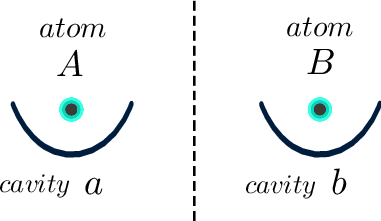}%
  \caption{{\footnotesize A schematic figure of the DJCM. In the left (right) partition
  there is the atom ``$A$" (``$B$") interacting with the cavity ``$a$" (``$b$"), respectively, and
  there is no interaction between the partition ``$Aa$" and ``$Bb$".}}\label{modelo}
% \end{center}
\end{figure}

We consider the cavities initially in the vacuum state
and some entanglement between the atoms. Consider the initial state
of the system as
\begin{eqnarray}\label{psi_0} %
|\psi_0\rangle&=&(\cos\alpha|10\rangle+\sin\alpha|01\rangle)\otimes|00\rangle
\ \ \ \ .
\end{eqnarray}
Because of the conservation of the number of excitations the time evolution can
be determined analytically and it reads
\begin{eqnarray}
|\psi_t\rangle&=&\nonumber
x_1(t)|10\rangle|00\rangle+x_2(t)|01\rangle|00\rangle+\\
\label{psi_t}&&+x_3(t)|00\rangle|10\rangle+x_4(t)|00\rangle|01\rangle
\ \ \ \ .
\end{eqnarray}

The coefficients will be given by the Schr\"odinger equation,
$i|\dot{\psi_t}\rangle=H|\psi_t\rangle$, plus the boundary
conditions $x_1(0)=\cos\alpha$, $x_2(0)=\sin\alpha$,
$x_3(0)=0$ and $x_4(0)=0$. They are
\begin{eqnarray}
  x_1(t) &=& \cos\alpha\cos(gt) \label{x1},\\
  x_2(t) &=& \sin\alpha\cos(gt) \label{x2},\\
  x_3(t) &=& -i\cos\alpha\sin(gt) \label{x3},\\
  x_4(t) &=& -i\sin\alpha\sin(gt) \label{x4}.
\end{eqnarray}
Consider also the initial state
\begin{eqnarray}\label{phi_0} %
|\phi_0\rangle&=&(\cos\alpha|11\rangle+\sin\alpha|00\rangle)\otimes|00\rangle
\ \ \ \ .
\end{eqnarray}
The same thing can be done to find the time evolution. We have
\begin{eqnarray}
|\phi_t\rangle&=&\nonumber
y_1(t)|11\rangle|00\rangle+y_2(t)|00\rangle|00\rangle+y_3(t)|10\rangle|01\rangle
+\\&&\label{phi_t}+y_4(t)|01\rangle|10\rangle
+y_5(t)|00\rangle|11\rangle, \ \ \ \
\end{eqnarray}
where
\begin{eqnarray}
  y_1(t) &=& e^{-i\omega t}\cos\alpha\cos^2(gt) \label{y1},\\
  y_2(t) &=& e^{i\omega t}\sin\alpha \label{y2},\\
  y_3(t) &=& -i\,e^{-i\omega t}\cos\alpha\sin(gt)\cos(gt) \label{y3},\\
  y_4(t) &=&  -i\,e^{-i\omega t}\cos\alpha\sin(gt)\cos(gt) \label{y4},\\
  y_5(t) &=&  -e^{-i\omega t}\cos\alpha\sin^2(gt)
  \label{y5} \ \ \ \ .
\end{eqnarray}
We can observe that, at time $t$ immediately after $t=0$, the
state (\ref{psi_t}) and (\ref{phi_t}) will develop entanglement among all
the partitions. However, we will consider the entanglement between
{\it qubits} ($A$, $B$, $a$ e $b$) and their relations. Thus, we will
use as entanglement quantifier the concurrence \cite{wootters2},
which is defined as
\begin{equation}\label{conc_def}
C=\max\Bigl[\,0\,,\sqrt{\lambda_1}-\sqrt{\lambda_2}-\sqrt{\lambda_3}-\sqrt{\lambda_4}\,\Bigr],
\end{equation}
where $\lambda_i$ are the eigenvalues, organized in a descending order, of the matrix
$\rho(\sigma_y\otimes\sigma_y)\rho^*(\sigma_y\otimes\sigma_y)$.

\section{Entanglement dynamics in the diagram $C_{ij}
\,\mathrm{x}\, C_{kl}$}
\label{Din_emara}

We can easily find the state $\rho_{ij}$ of two {\it qubits} taking a
partial trace over the remaining subsystem. 
We next determine all $C_{ij}$.

\subsection{ For the initial state $|\psi_0\rangle$}

In this case we obtain
\begin{eqnarray}
  C_{AB} &=& |\sin 2\alpha|\cos^2(gt) \label{c1},\\
  C_{ab} &=& |\sin 2\alpha|\sin^2(gt) \label{c2},\\
  C_{Aa} &=& \cos^2\alpha|\sin(2gt)| \label{c3},\\
  C_{Ab} &=& |\sin 2\alpha\sin(gt)\cos(gt)| \label{c4},\\
  C_{aB} &=& |\sin 2\alpha\sin(gt)\cos(gt)| \label{c5},\\
  C_{Bb} &=& \sin^2\alpha|\sin(2gt)| \ \ \ \ .\label{c6}
\end{eqnarray}

We analyze the geometric structure of entanglement dynamics in a
diagram $C_{ij} \,\mathrm{x}\, C_{kl}$. In order to do this, observe that we
can sum eq.(\ref{c1}) with eq.(\ref{c2}) and we have

\begin{equation}\label{c11}
C_{AB}+C_{ab} = C_0,
\end{equation}
\begin{figure}[h]

\centering
  \includegraphics[scale=0.300,angle=-90]{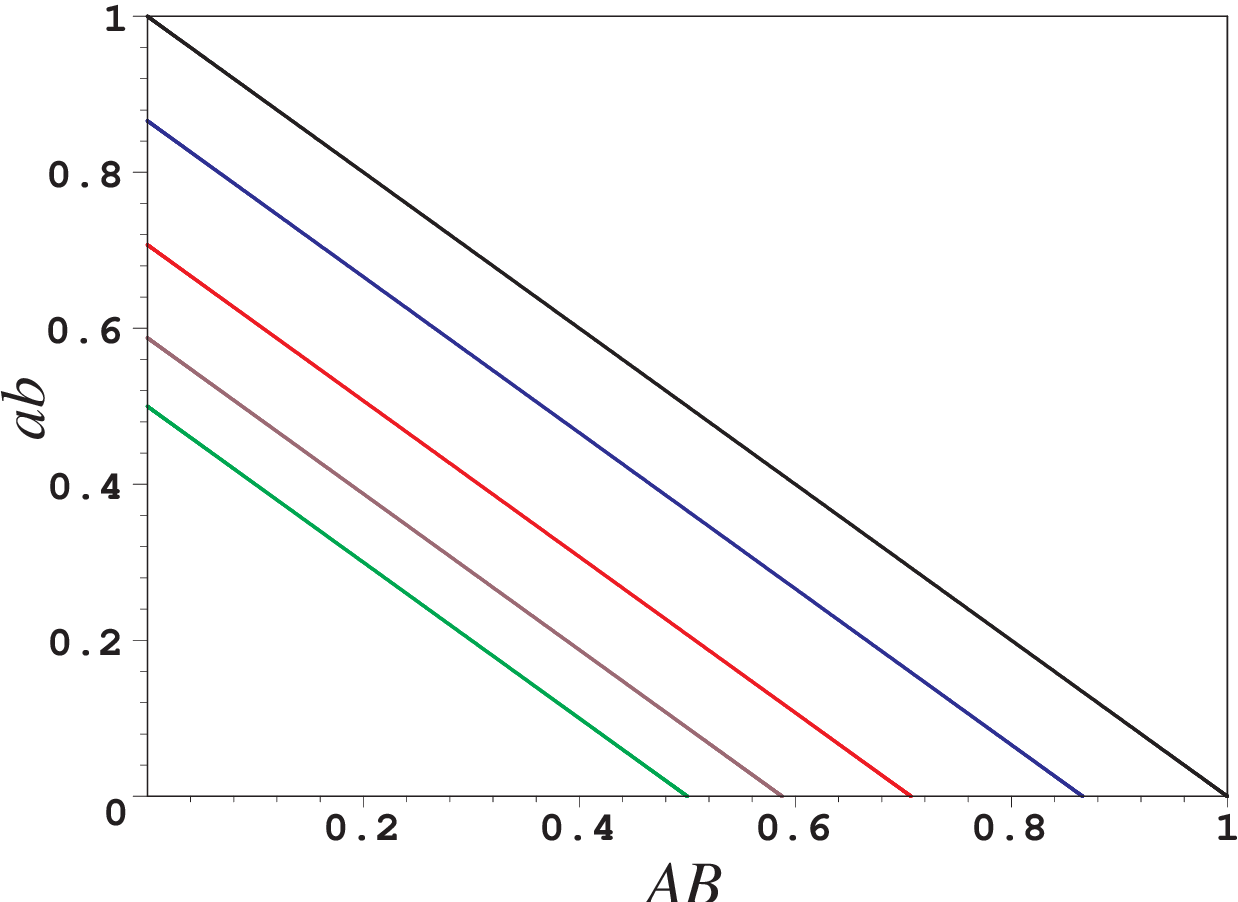}%\\
  \caption{{\footnotesize Graphic of the straight lines $C_{AB} \,\mathrm{x}\, C_{ab}$
  with $\alpha=\pi/4, \pi/6, \pi/8, \pi/10 \ \mathrm{and} \ \pi/12$ for the colors black, blue, red, brown and green, respectively.}}\label{fig01}
\end{figure}%
where $C_0=|\sin 2\alpha|$ is the initial concurrence between the
atoms $A$ and $B$. We notice that this equation defines a straight line in
a diagram $C_{AB} \,\mathrm{x}\, C_{ab}$ (see figure \ref{fig01}).
The lines in equation (\ref{c11}), when $\alpha\in(0,\pi/2)$, fill
the triangle formed by the axis $C_{AB}$, $C_{ab}$ and
$C_{AB}+C_{ab} =1$. In addition, we notice that equations (\ref{c4}) and
(\ref{c5}) satisfy

\begin{equation}\label{c12}
C_{Ab}=C_{aB}.
\end{equation}
This shows a symmetry between the cavity of one of the systems and the atom of the other.
We proceed dividing (\ref{c3}) by (\ref{c6}) and we easily find

\begin{equation}\label{c13}
C_{Aa}=\dfrac{\cos^2\alpha}{\sin^2\alpha}C_{Bb}, \ \ \ \
\end{equation}
\begin{figure}[h]
%\vspace{110pt}
%\hspace{-200pt}
\centering
  \includegraphics[scale=0.300,angle=-90]{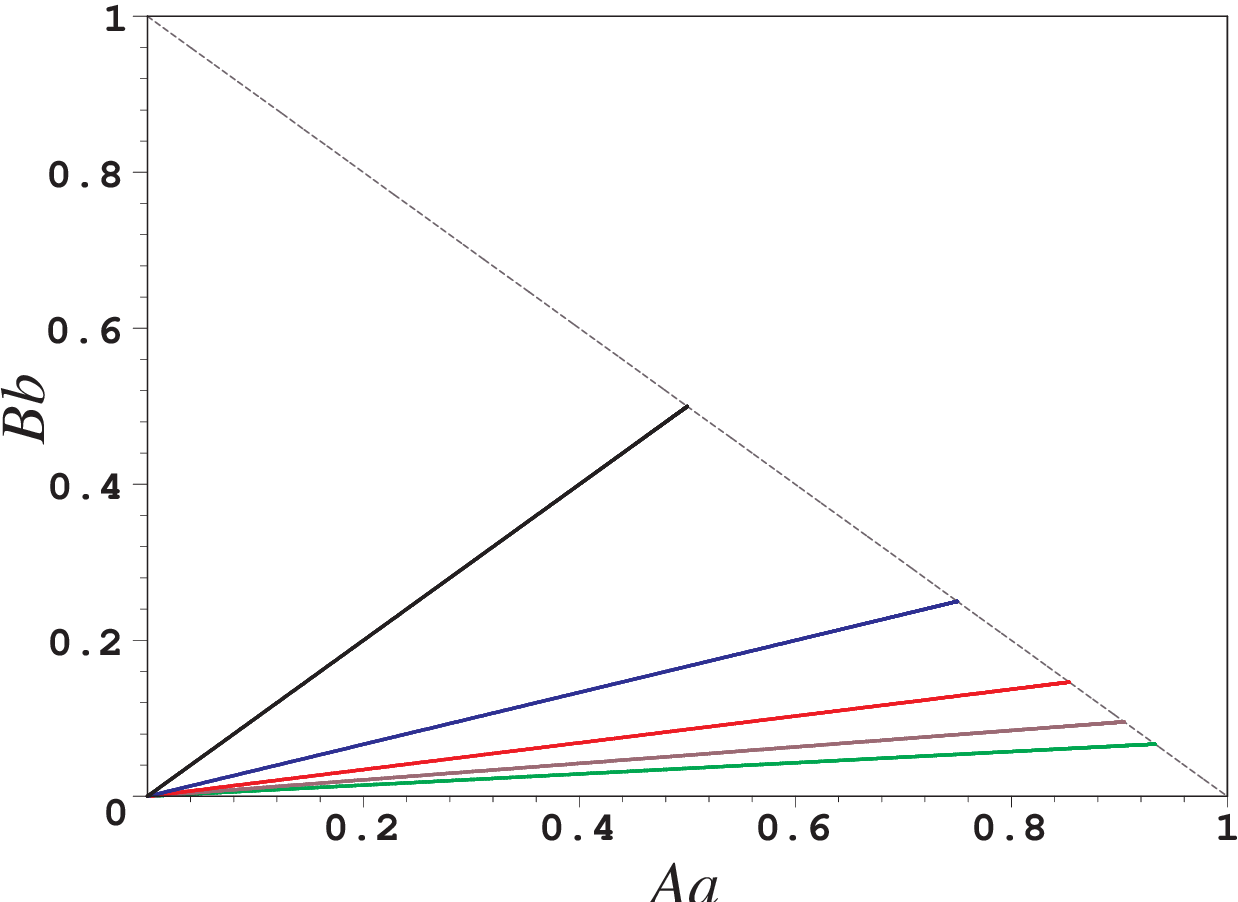}%\\
  \caption{{\footnotesize Graphic of the straight line $C_{Aa} \,\mathrm{x}\, C_{Bb}$
  with $\alpha=\pi/4, \pi/6, \pi/8, \pi/10 \ \mathrm{and} \ \pi/12$ for the colors black,
 blue, red, brown and green, respectively. The slim violet curve is the straight line $C_{Aa}+C_{Bb}=1$.}}\label{fig02}
\end{figure}
which is a straight line in the diagram $C_{Aa} \,\mathrm{x}\,
C_{Bb}$. In the interval $0<\alpha<\pi/2$, the lines (\ref{c13}) are
limited in the region between the lines $C_{Aa}=0$, $C_{Bb}=0$ and
$C_{Aa}+C_{Bb}=1$. Equations (\ref{c11} --
\ref{c13}) define a straight line in their respective diagram
$C_{ij} \,\mathrm{x}\, C_{kl}$. The line $C_{Aa}+C_{Bb}=1$, which represents a conservation of entanglement, is a
superior limit in all cases. Using the same procedure, and some simplifications,
we find other conics (ellipses, circumferences and straight lines)
which we organize as follows:

\subsubsection{ Concurrence between atoms (or cavities) {\it versus}
concurrence between one of the atoms and its cavity:} \label{s3.1.1}
\begin{itemize}
{\it
    \item [a)] $C_{AB(ab)} \,\mathrm{x}\, C_{Bb}$:
    
\begin{equation}
  \dfrac{\bigl(C_{AB(ab)}-C_{0}^{}/2\bigr)^2}{C_{0}^{2}/4}
  +\dfrac{C_{Bb}^2}{\sin^{4}\alpha} = 1 \label{elipse_AB(ab)_Bb}
\end{equation}

\begin{figure}[h]

\centering
  \includegraphics[scale=0.300,angle=-90]{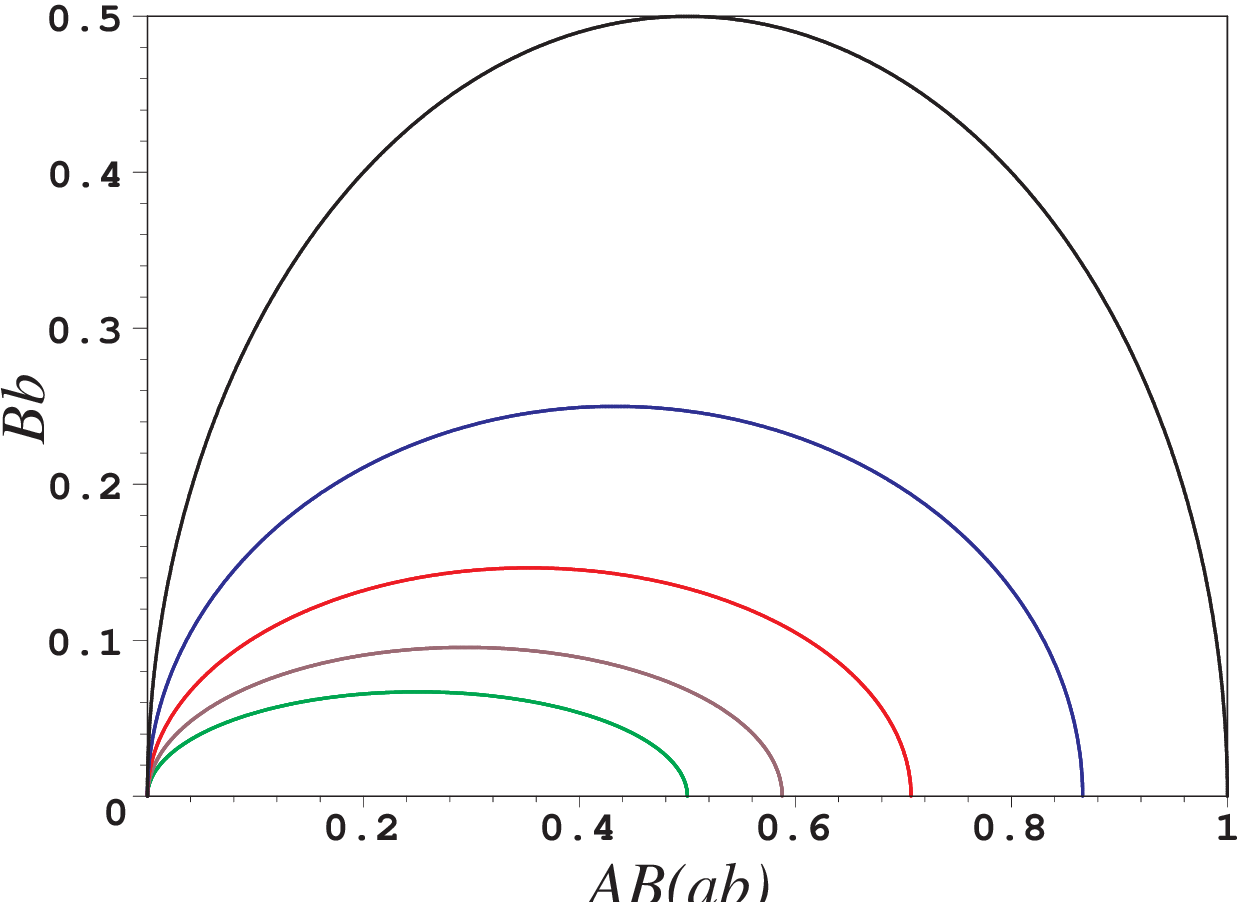}%\\
  \caption{{\footnotesize Graphic of the semi--ellipse $C_{AB(ab)} \,\mathrm{x}\, C_{Bb}$
  with $\alpha=\pi/4, \pi/6, \pi/8, \pi/10 \ \mathrm{and} \ \pi/12$ for the colors black, blue, red, brown and green, respectively.}}\label{fig2}
\end{figure}

    \item [ b)] $C_{AB(ab)} \,\mathrm{x}\, C_{Aa}$:
    
\begin{equation}
  \dfrac{\bigl(C_{AB(ab)}-C_{0}^{}/2\bigr)^2}{C_{0}^{2}/4}
  +\dfrac{C_{Aa}^2}{\cos^{4}\alpha} = 1 \label{elipse_AB(ab)_Aa}
\end{equation}

\begin{figure}[h]

\centering
  \includegraphics[scale=0.300,angle=-90]{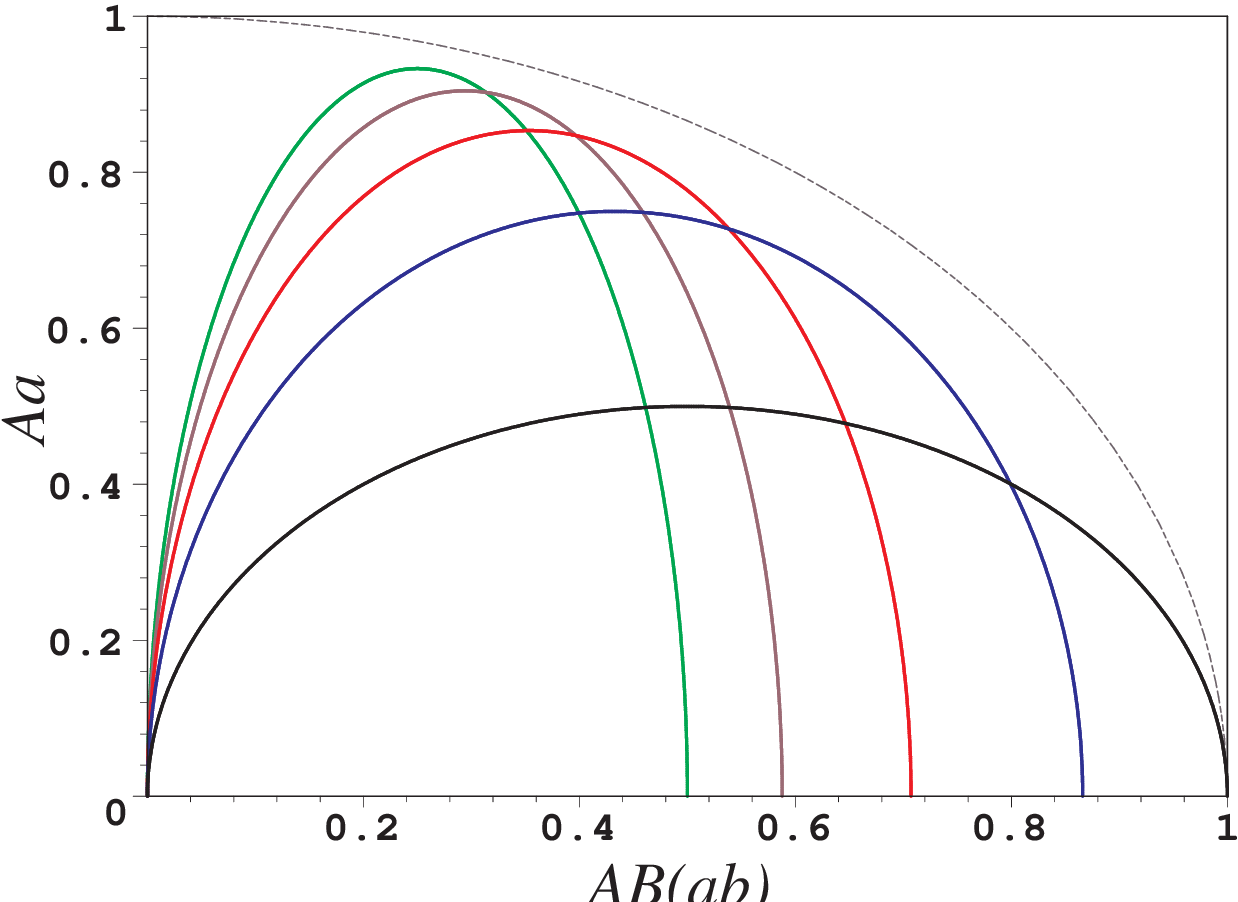}%\\
  \caption{{\footnotesize Graphic of the semi--ellipse $C_{AB(ab)} \,\mathrm{x}\, C_{Aa}$
  with $\alpha=\pi/4, \pi/6, \pi/8, \pi/10 \ \mathrm{and} \ \pi/12$ for the colors black, blue,
  red, brown and green, respectively.
  The slim violet curve is the semi circumference $C_{AB(ab)}^{2}+C_{Aa}^{2}=1$.}}\label{fig3}
\end{figure}

}
\end{itemize}

\subsubsection{ Concurrence between atoms (or cavities) {\it versus}
concurrence between one of the atoms and the cavity which does not contain it:}
\begin{equation}
  \bigl(C_{AB(ab)}-C_{0}^{}/2\bigr)^2
  +\bigl(C_{aB(Ab)}\bigr)^2 = C_{0}^{2}/4 \label{elipse_AB(ab)_aB(Ab)}
\end{equation}
\begin{figure}[h]
\centering
  \includegraphics[scale=0.300,angle=-90]{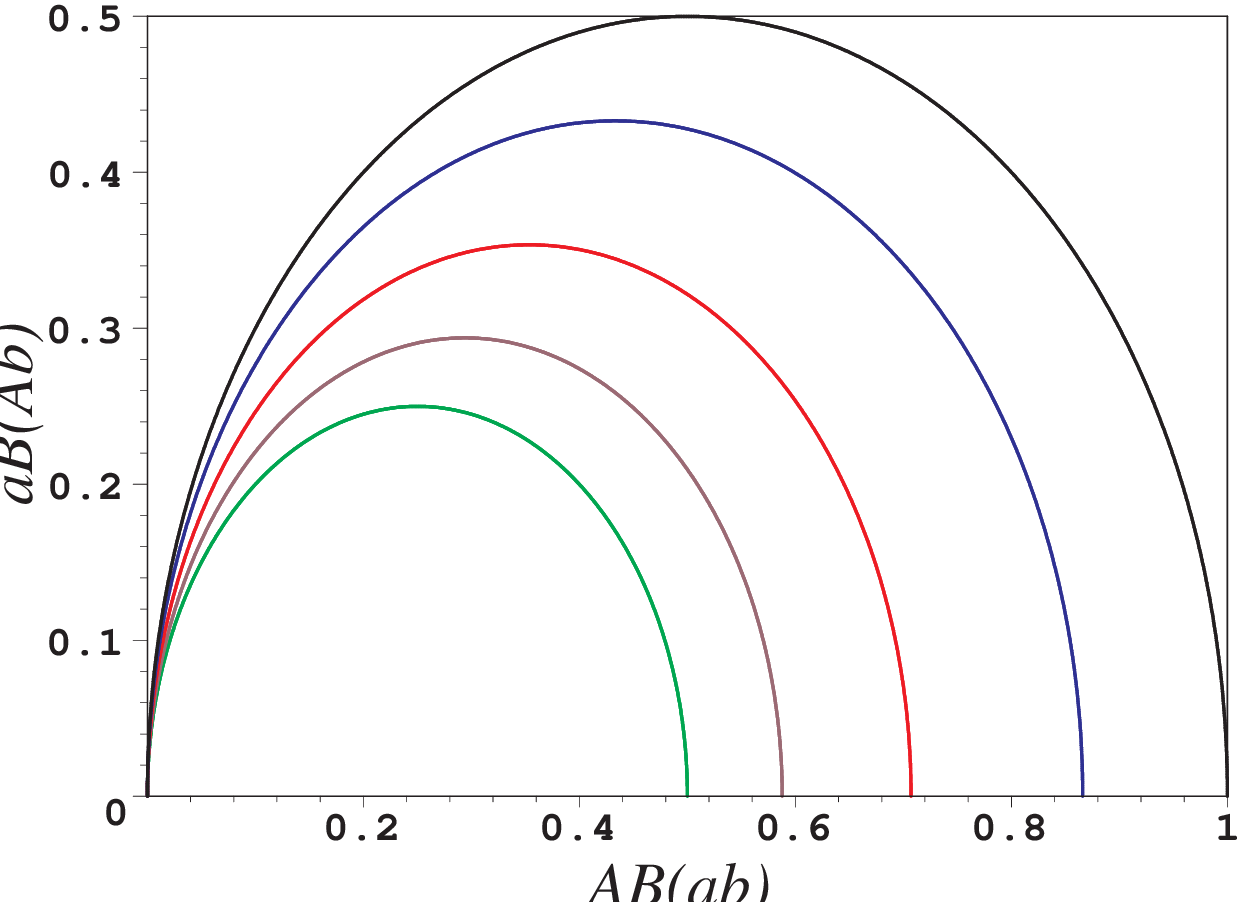}%\\
  \caption{{\footnotesize Graphic of the semi circumference $C_{AB(ab)} \,\mathrm{x}\, C_{aB(Ab)}$
  with $\alpha=\pi/4, \pi/6, \pi/8, \pi/10 \ \mathrm{and} \ \pi/12$ for the colors black, blue,
  red, brown and green, respectively.}}\label{fig4}
\end{figure}

\subsubsection{ Concurrence between one of the atoms and the cavity which does not contain it
{\it versus} concurrence between one of the atoms and its cavity:} \label{s3.1.3}

\begin{itemize}
{\it
    \item [a)] $C_{aB(Ab)} \,\mathrm{x}\, C_{Aa}$:
    
\begin{equation}\label{elipse_aB(Ab)_Aa}
C_{aB(Ab)}=\dfrac{|\sin2\alpha|}{2\cos^2\alpha}C_{Aa}
\end{equation}
\begin{figure}[h]

\centering
  \includegraphics[scale=0.300,angle=-90]{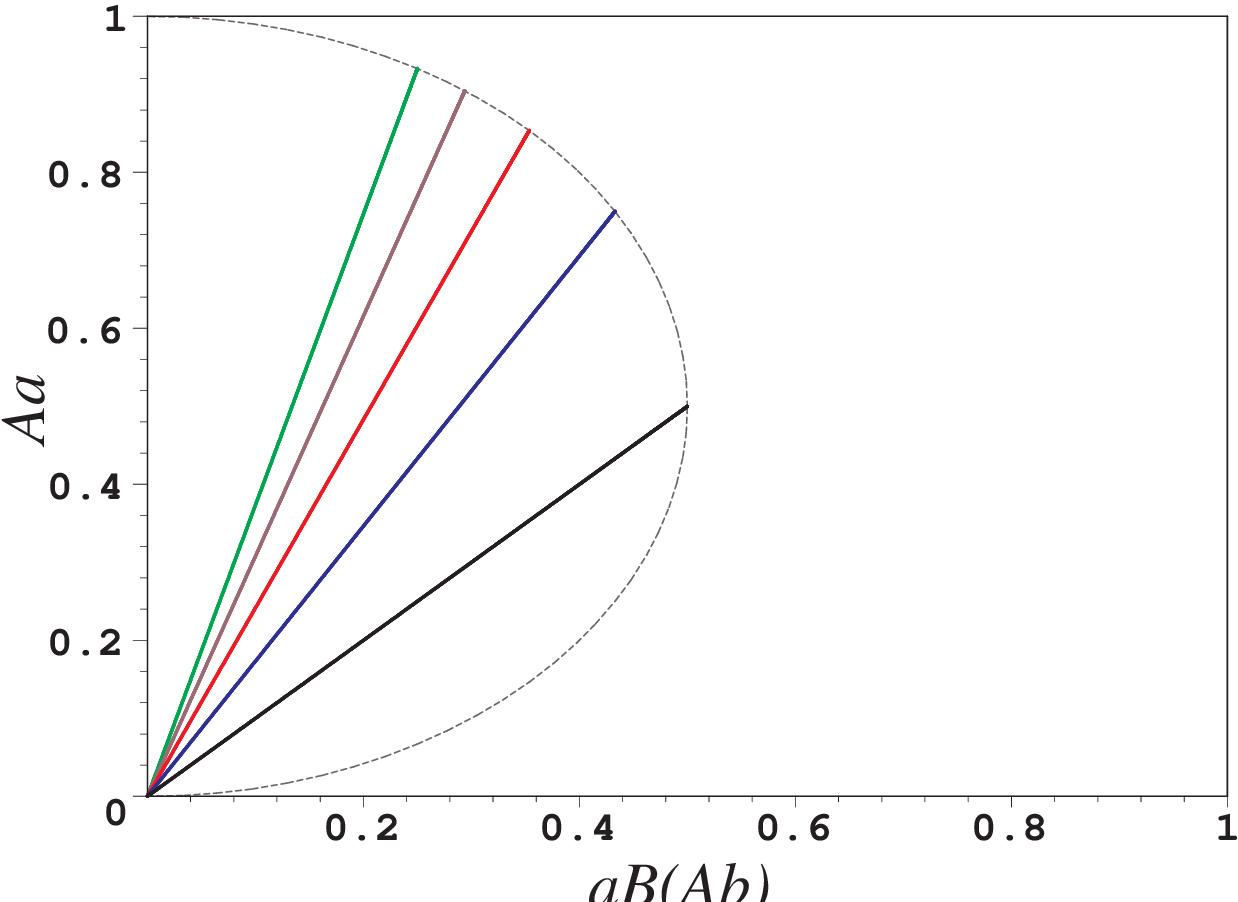}%\\
  \caption{{\footnotesize Graphic of the straight line $C_{aB(Ab)} \,\mathrm{x}\, C_{Aa}$
  with $\alpha=\pi/4, \pi/6, \pi/8, \pi/10 \ \mathrm{and} \ \pi/12$ for the colors black, blue,
  red, brown and green, respectively.
  The slim violet curve is the semi circumference $(2C_{Aa}-1)^2+(2C_{aB(Ab)})^2=1$.}}\label{fig5}
\end{figure}

    \item [ b)] $C_{Ab(aB)} \,\mathrm{x}\, C_{Bb}$:
    
\begin{equation}
C_{Ab(aB)}=\dfrac{|\sin2\alpha|}{2\sin^2\alpha}C_{Bb}
\label{elipse_Ab(aB)_Bb}
\end{equation}

\begin{figure}[h]
\centering
  \includegraphics[scale=0.300,angle=-90]{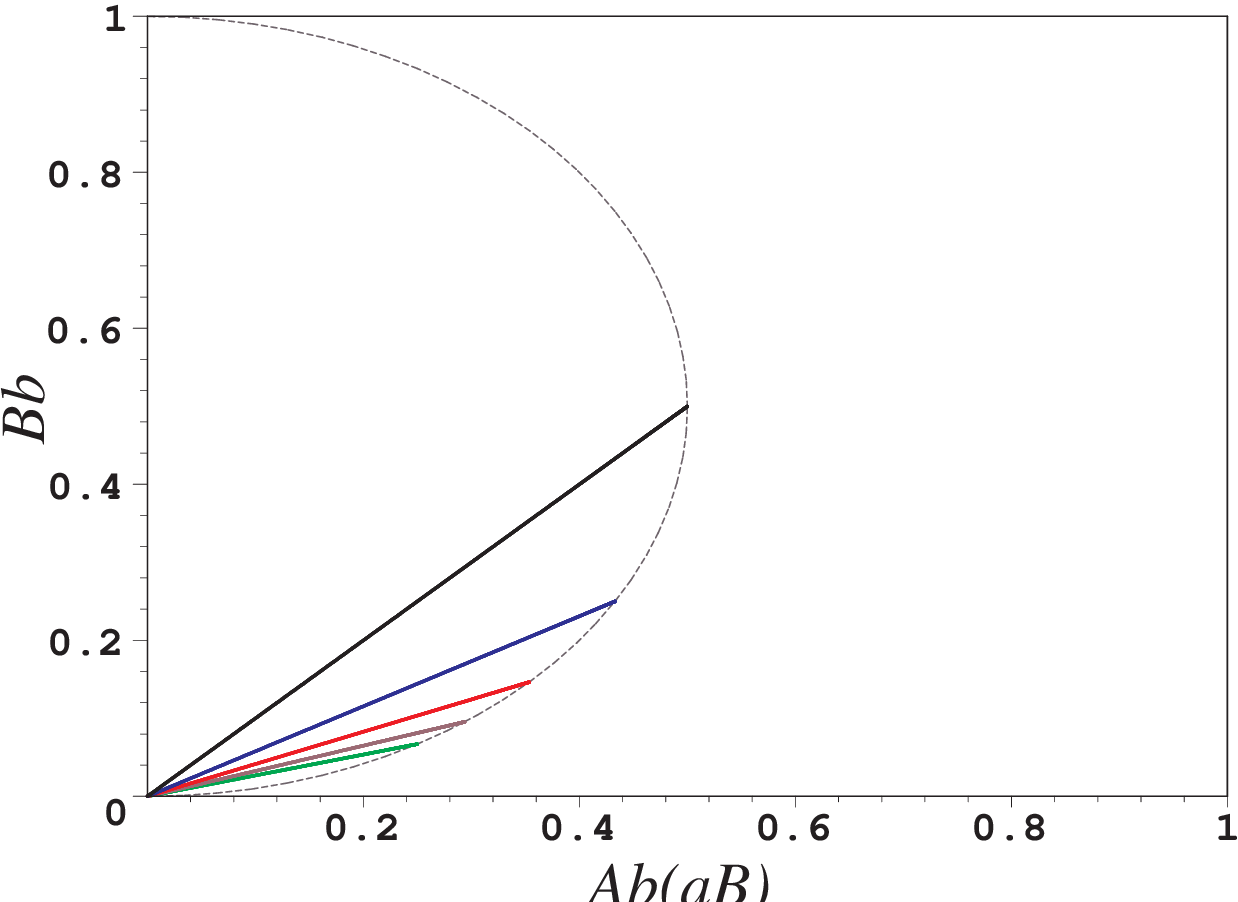}%\\
  \caption{{\footnotesize Graphic of the straight line $C_{Ab(aB)} \,\mathrm{x}\, C_{Bb}$
  with $\alpha=\pi/4, \pi/6, \pi/8, \pi/10 \ \mathrm{and} \ \pi/12$ for the colors black, blue,
  red, brown and green, respectively.
  The slim violet curve is the semi circumference $(2C_{Bb}-1)^2+(2C_{Ab(aB)})^2=1$.}}\label{fig6}
\end{figure}} 

\end{itemize}

In order to interpret the expressions (\ref{elipse_AB(ab)_Bb} --
\ref{elipse_Ab(aB)_Bb}) and their respective figures (\ref{fig2} --
\ref{fig6}), it becomes instructive to use the {\it predictability},

\begin{equation}\label{predi}
\mathcal{P}_0=\Bigl|\mathrm{tr}\bigl(
\sigma_{z}^{A}\rho_0\bigr)\Bigr| \ \ \ \ \ .
\end{equation}

We use predictability because, unlike concurrence, it
is measurable (the module of the mean value of an observable), local and
it is related to the concurrence \cite{bergou}.
For $\rho_0=|\psi_0\rangle\langle\psi_0|$ we have
$\mathcal{P}_0=|\cos(2\alpha)|$, and it is clear that
$C_{0}^{2}+\mathcal{P}_{0}^{2}=1$. Observe that when
$\mathcal{P}_0=0$ the excitation will be equally distributed
between the partitions $Aa$ and $Bb$, it will not be localized and the
initial entanglement will be maximum between $A$ and $B$. On the other hand, if
$\mathcal{P}_0=1$ the atoms will not be initially entangled
and the information if the excitation
will be in partition $Aa$ or $Bb$ will not be available. However, we can assure that the
excitation will be in the partition $Aa$ or in the partition $Bb$. When
$0<\mathcal{P}_0<1$, all we know is that the excitation has a larger probability to be in
one of the partitions.

The eccentricity of the
ellipses (\ref{elipse_AB(ab)_Bb}) and (\ref{elipse_AB(ab)_Aa}) can be written as
a function of the {\it predictability}
\begin{equation}\label{excen}
\epsilon=\sqrt{\dfrac{2\mathcal{P}_0}{1+\mathcal{P}_0}} \ \ \ \ .
\end{equation}
We can determine also the distance $f$ of the focus to the center of
each ellipse. For the ellipse (\ref{elipse_AB(ab)_Bb}) the distance of
the focus $f^{(a)}$ to its center will be
\begin{equation}\label{foco_a}
f^{(a)}_{\lessgtr}=\sqrt{\dfrac{\mathcal{P}_0(1\mp\mathcal{P}_0)}{2}},
\end{equation}
where $f^{(a)}_{<}$ is the focus if $0<\alpha<\pi/4$ and
$f^{(a)}_{>}$ is the focus if $\pi/4<\alpha<\pi/2$. The ellipse
(\ref{elipse_AB(ab)_Aa}) will have the focus $f^{(b)}$ as being
\begin{equation}\label{foco_b}
f^{(b)}_{\gtrless}=\sqrt{\dfrac{\mathcal{P}_0(1\pm\mathcal{P}_0)}{2}},
\end{equation}
where $f^{(b)}_{>}$ is the focus if $0<\alpha<\pi/4$ and
$f^{(b)}_{<}$ is the focus if $\pi/4<\alpha<\pi/2$, i. e. the
opposite case of (\ref{foco_a}). This happens because the
entanglement of the partition $Aa$ ($Bb$) is generated by the
JCM evolution and not by the initial source of entanglement
contained in $AB$. The entanglement generated by the JCM depends on
the ``quantity" of excitation that will be shared between the
respective atom--field. Thus, when $0<\alpha<\pi/4$, the excitation,
in the state represented by (\ref{psi_t}), will be more likely to be
found in the partition $Aa$. Then, the entanglement generated by the
JCM in the partition $Aa$ will be larger than $Bb$. This is
represented in figure \ref{fig3}, where $C_{Aa}$ reaches larger
values than $0.5$ if $0<\alpha<\pi/4$. In this case, the
entanglement in the partition $Bb$ has values below 0.5, as we can
observe in figure \ref{fig2}. The same analysis is valid in the
opposite case, where $\pi/4<\alpha<\pi/2$. On the other hand, if
$\alpha\in(0,\pi/2)$, the eccentricity of the ellipses
(\ref{elipse_AB(ab)_Bb}) and (\ref{elipse_AB(ab)_Aa}) are identical,
as shown in (\ref{excen}), but the focuses $f^{a}$ and $f^{b}$ do
not have the same value and do not necessarily lie in the same axis,
except for the case $\alpha=\pi/4$ when we have circumferences in
both cases. For example, if $\alpha=\pi/6$ we have
$f^{b}=\sqrt{3\,}f^{a}$, where $f^a$ ($f^b$) is over the horizontal
(vertical) axis, respectively.

In section \ref{s3.1.3}, equation (\ref{elipse_AB(ab)_aB(Ab)}) represents
semi circumferences with radius $C_{0}^{}/2$. When  
$\alpha=\pi/4$ ($C_0=1$) the limiting curve is obtained. More
generally, we can say that a curve defined in its respective diagram
$C_{ij} \,\mathrm{x}\, C_{kl}$ is always limited by the
semi circumference $C_{ij}^{2} + C_{kl}^{2}=C_{0}^{2}$.

The sequential cases, represented by equations
(\ref{elipse_aB(Ab)_Aa}) and (\ref{elipse_Ab(aB)_Bb}), are straight
lines with angular coefficient dependent on the initial
entanglement. As we did previously, we can write the angular
coefficient as functions of the {\it predictability}.
Equation(\ref{elipse_aB(Ab)_Aa}) has angular coefficient $m^{(a)}$
given by
\begin{equation}\label{m_a}
m^{(a)}_{\lessgtr}=\biggl[\dfrac{1-\mathcal{P}_{0}}{1+\mathcal{P}_0}\biggr]^{\pm
1/2} \ \ \ \ ,
\end{equation}
where $m^{(a)}_{<}$ is the coefficient when $0<\alpha<\pi/4$ and
$m^{(a)}_{>}$ is the coefficient if $\pi/4<\alpha<\pi/2$. Straight
lines of equation (\ref{elipse_Ab(aB)_Bb}) have angular coefficient
\begin{equation}\label{m_b}
m^{(b)}_{\gtrless}=\biggl[\dfrac{1-\mathcal{P}_{0}}{1+\mathcal{P}_0}\biggr]^{\pm
1/2} \ \ \ \ ,
\end{equation}
where, as in the previous case, $m^{(b)}_{<}$ is the coefficient if
$0<\alpha<\pi/4$ and $m^{(b)}_{>}$ is the coefficient if
$\pi/4<\alpha<\pi/2$. The opposite occurs for (\ref{m_a}). This
effect is also due to the entanglement given by the JCM as we
already discussed previously in the ellipse equations
(\ref{elipse_AB(ab)_Bb}) and (\ref{elipse_AB(ab)_Aa}).

\subsection{For the initial state $|\phi_0\rangle$}

Let us now consider the physical system whose initial state is given
by equation (\ref{phi_0}). After a time interval $t$, the state 
of the system will be (\ref{phi_t}). In an analogous way as before we
determine the concurrences of each pair of {\it qubits}. Those are
\begin{eqnarray}
  {C}_{AB} &=& \max\bigl[0,C_0\cos^{2}(gt)-\gamma_t\bigr] \label{cc1},\\
  {C}_{ab} &=& \max\bigl[0,C_0\sin^{2}(gt)-\gamma_t\bigr] \label{cc2},\\
  {C}_{Aa} &=& \cos^{2}(\alpha)|\sin^{}(2gt)| \label{cc3},\\
  {C}_{Ab} &=& \max\bigl[0,\dfrac{1}{2}C_0|\sin^{}(2gt)|-\gamma_t\bigr] \label{cc4},\\
  {C}_{aB} &=& \max\bigl[0,\dfrac{1}{2}C_0|\sin^{}(2gt)|-\gamma_t\bigr] \label{cc5},\\
  {C}_{Bb} &=& \cos^{2}(\alpha)|\sin^{}(2gt)| \label{cc6} \
  \ \ \ \ ,
\end{eqnarray}
where $\gamma_t=\dfrac{1}{2}\cos^{2}(\alpha)\sin^{2}(2gt)$. Observe
that if $0<\alpha<\pi/4$, we have entanglement sudden death
\cite{morte} or entanglement sudden birth \cite{nascimento}. 

In this case, we have some interesting situations due to the
symmetry of the system. Notice that the partition $Aa$ and $Bb$ will
have the same value of {\it predictability}. Thus, the dynamical
entanglement supplied by the JCM  to $Aa$ or $Bb$ is the same.
Observe that ${C}_{Aa}= {C}_{Bb}$. This would not be true if the
coupling constant of each JCM was different. Due to that same
symmetry we also have ${C}_{Ab}= {C}_{Ba}$. Those relations define
straight lines (like the case of equation (\ref{c12})) in their
respective diagrams. The other diagrams $C_{ij} \,\mathrm{x}\,
C_{kl}$, however, are not so simple. That is because the initial
state (\ref{phi_0}) contains the eigenstate $|00\rangle \otimes |00\rangle$
of the Hamiltonian (\ref{H}), which does not contributes for the entanglement
generated by the JCM, i. e. the time evolution of that eigenstate only adds a global
phase to it (see eq. (\ref{y2})). On the other hand, if the initial state is (\ref{psi_0}),
both the states $|10\rangle \otimes |00\rangle$ and $|01\rangle \otimes |00\rangle$ contribute
for the entanglement generated by the JCM in a form of senoidal functions of time in the amplitudes
of the state (\ref{psi_t}) and that is why we obtain conics when we make parametric plots 
of concurrences.

\begin{figure}[h]

\centering
  \includegraphics[scale=0.300,angle=-90]{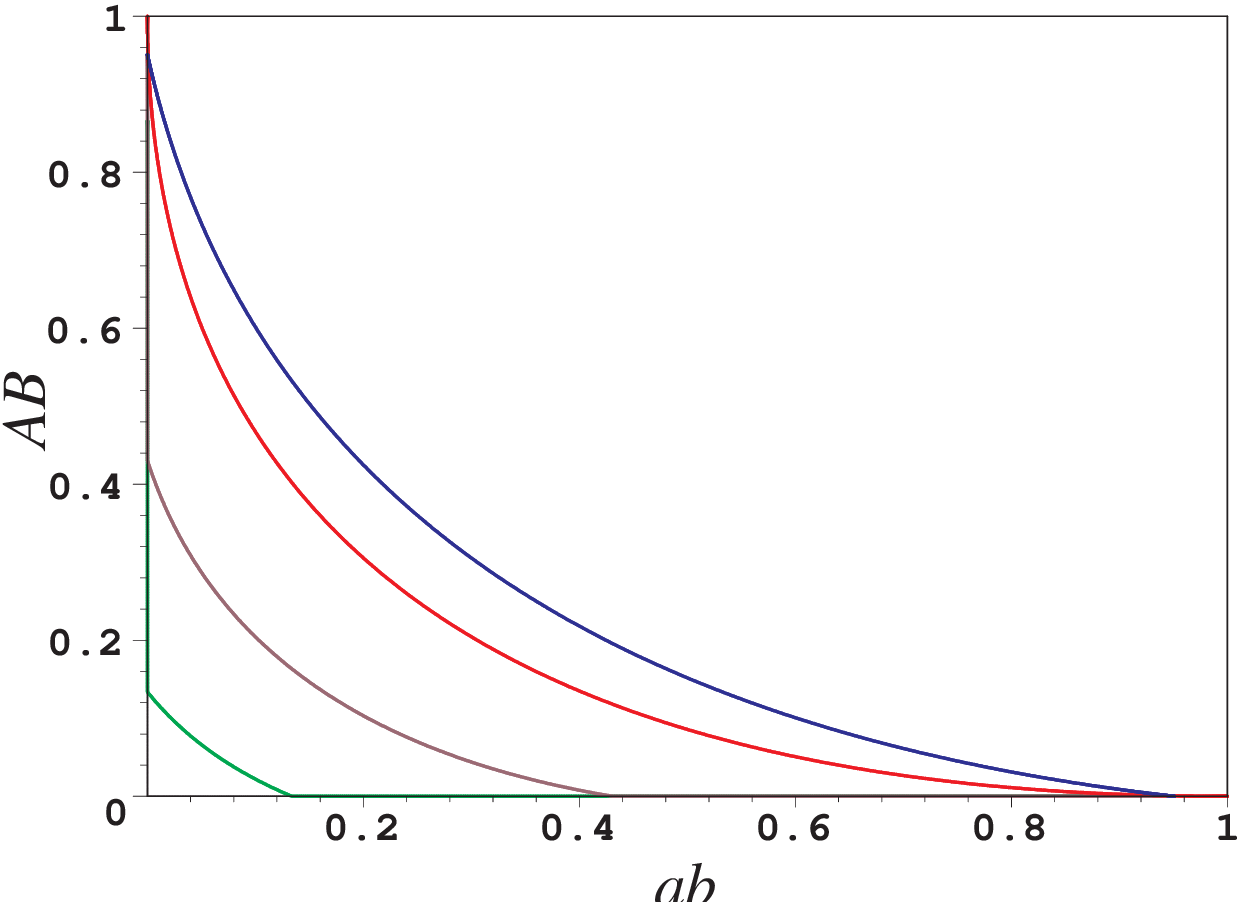}%\\
  \caption{{\footnotesize Graphic of the parabola ${C}_{ab} \,\mathrm{x}\, {C}_{AB}$
  with $\alpha=3\pi/10, \pi/4, \pi/5\ \mathrm{and} \ \pi/6 $ for the colors blue,
  red, brown and green, respectively.
}}\label{fig7}
\end{figure}

The next case is in the diagram $C_{ab} \,\mathrm{x}\, C_{AB}$.
Consider an instant of time when the concurrences ${C}_{AB}$ and
${C}_{ab}$ are different from zero at the same time. Then, we can
write ${C}_{AB}=C_0\cos^{2}(gt)-\gamma_t$ and
${C}_{ab}=C_0\sin^{2}(gt)-\gamma_t$. Notice that using simple
algebra we can write
$[C_0-({C}_{AB}+{C}_{ab})]/\cos^{2}(\alpha)=\sin^2(2gt)$ and
$({C}_{AB}-{C}_{ab})^2/C_{0}^{2}=\cos^2(2gt)$. Summing both we have
\begin{equation}\label{parabola}
\dfrac{\bigl({C}_{AB}-{C}_{ab}\bigr)^2}{C_{0}^{2}}+\dfrac{C_0
-\bigl({C}_{AB}+{C}_{ab}\bigr)}{\cos^{2}(\alpha)}=1 \ \ \ \ .
\end{equation}
This is a parabola with symmetry axis at $45^{\mathrm{o}}$ of the
horizontal axis $({C}_{ab})$. On this axis the vertex $v$ is
localized at point $v_{\lessgtr}=\{0,C_{0}-(1\pm\mathcal{P}_0)/2\}$
and the focus $f$ at $f_{\lessgtr}=\{0,C_{0}\mp\mathcal{P}_0\}$,
where the index $<$ ($>$) refers to $0<\alpha<\pi/4$
($\pi/4<\alpha<\pi/2$), respectively. Because of the entanglement
sudden death in the partitions $AB$ and $ab$ whenever
$0<\alpha<\pi/4$, there will only be a segment of the parabola in
the diagram ${C}_{ab} \,\mathrm{x}\, {C}_{AB}$ if the vertex $v$
admits positive values on the axis of the parabola. On the other
hand, when the vertex is the origin or admits negative values, we
will only have the straight line ${C}_{AB}=0$ or ${C}_{ab}=0$
(observe figure \ref{fig7} for illustration).

\begin{figure}[h]
\centering
  \includegraphics[scale=0.300,angle=-90]{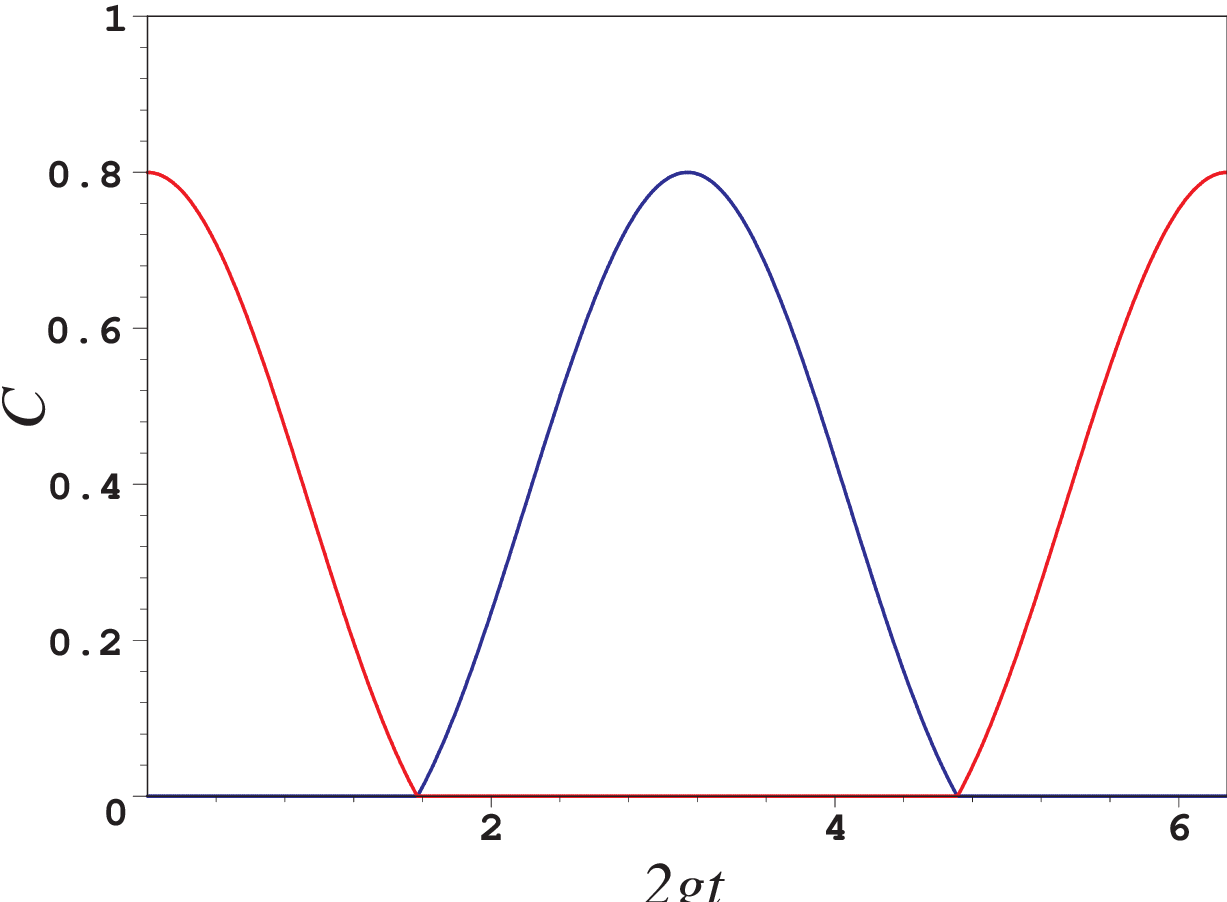}%\\
  \caption{{\footnotesize In red (blue) the graphic of ${C}_{AB}$
  (${C}_{ab}$), respectively, for $\alpha=\arctan(1/2)$.}}\label{fig8}
\end{figure}

If $\alpha=\arctan(1/2)$ we have $v=\{0,0\}$ and when the entanglement in one of the partitions disappears
the entanglement of another one resurges, as we see in figure \ref{fig8}.

\begin{figure}[h]

\centering
  \includegraphics[scale=0.300,angle=-90]{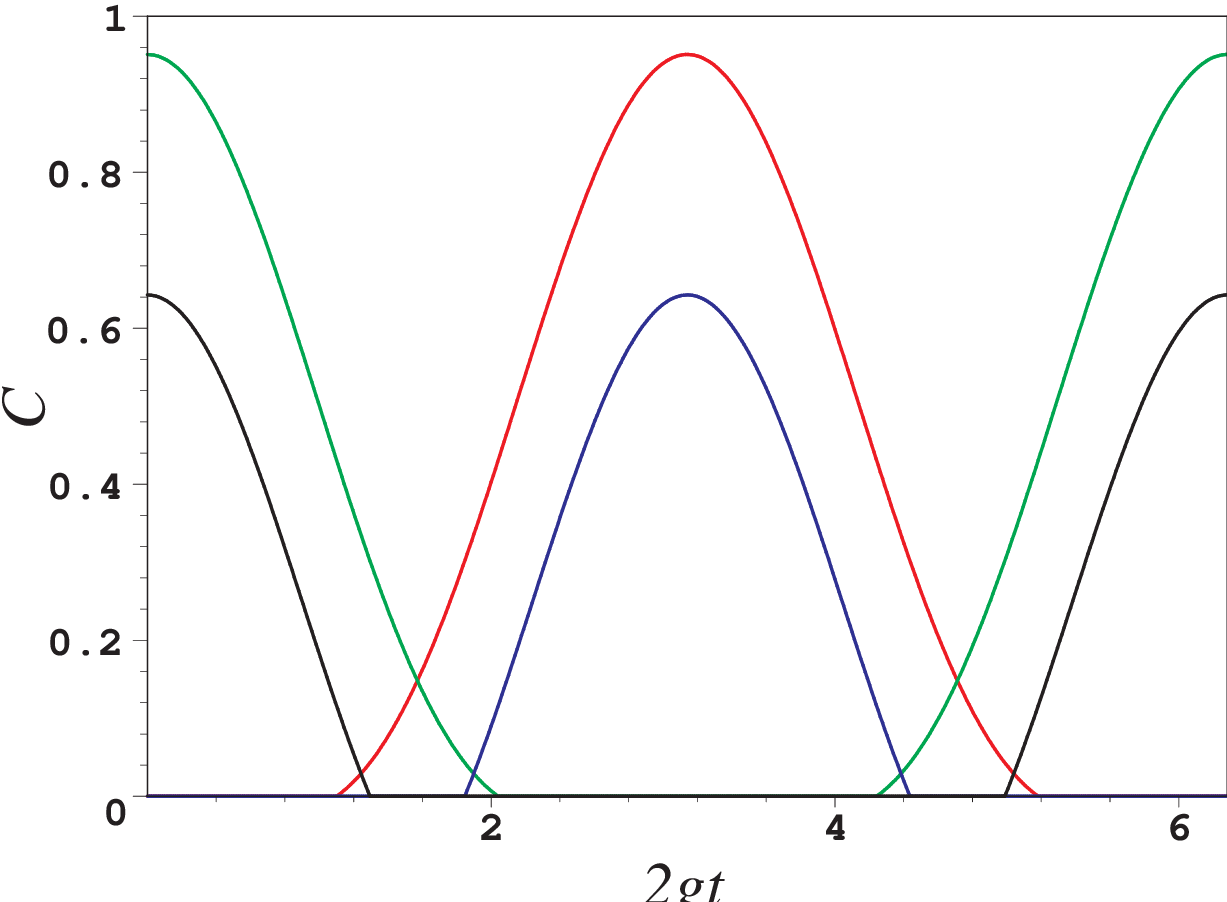}%\\
  \caption{{\footnotesize Graphic of ${C}_{AB}$
  (${C}_{ab}$) in black (blue) when $\alpha=\pi/9<\arctan(1/2)$, respectively.
	Graphic of ${C}_{AB}$ (${C}_{ab}$) in green (red) when $\alpha=\pi/5>\arctan(1/2)$,
 respectively.}}\label{fig9}
\end{figure}

Following the same reasoning, it is clear that if
$0<\alpha<\arctan(1/2)$ (or $\arctan(1/2)<\alpha<\pi/2$) the
entanglement in $AB$ disappears before (or after) it appears in
$ab$, respectively (this dynamics is depicted in figure \ref{fig9}).

We keep seeking for more relations. Using the sum of equations (\ref{cc1}) and (\ref{cc2}), squaring them and adding to
equations (\ref{cc3}) or (\ref{cc6}) squared, we get the following ellipse
with expression
\begin{equation}\label{elipse_phi}
\dfrac{\bigl({C}_{AB}-{C}_{ab}\bigr)^2}{C_{0}^{2}}+\dfrac{{C}^{2}_{Aa(Bb)}}{\cos^4\alpha}=1.
\end{equation}
\begin{figure}[h]

\centering
  \includegraphics[scale=0.300,angle=-90]{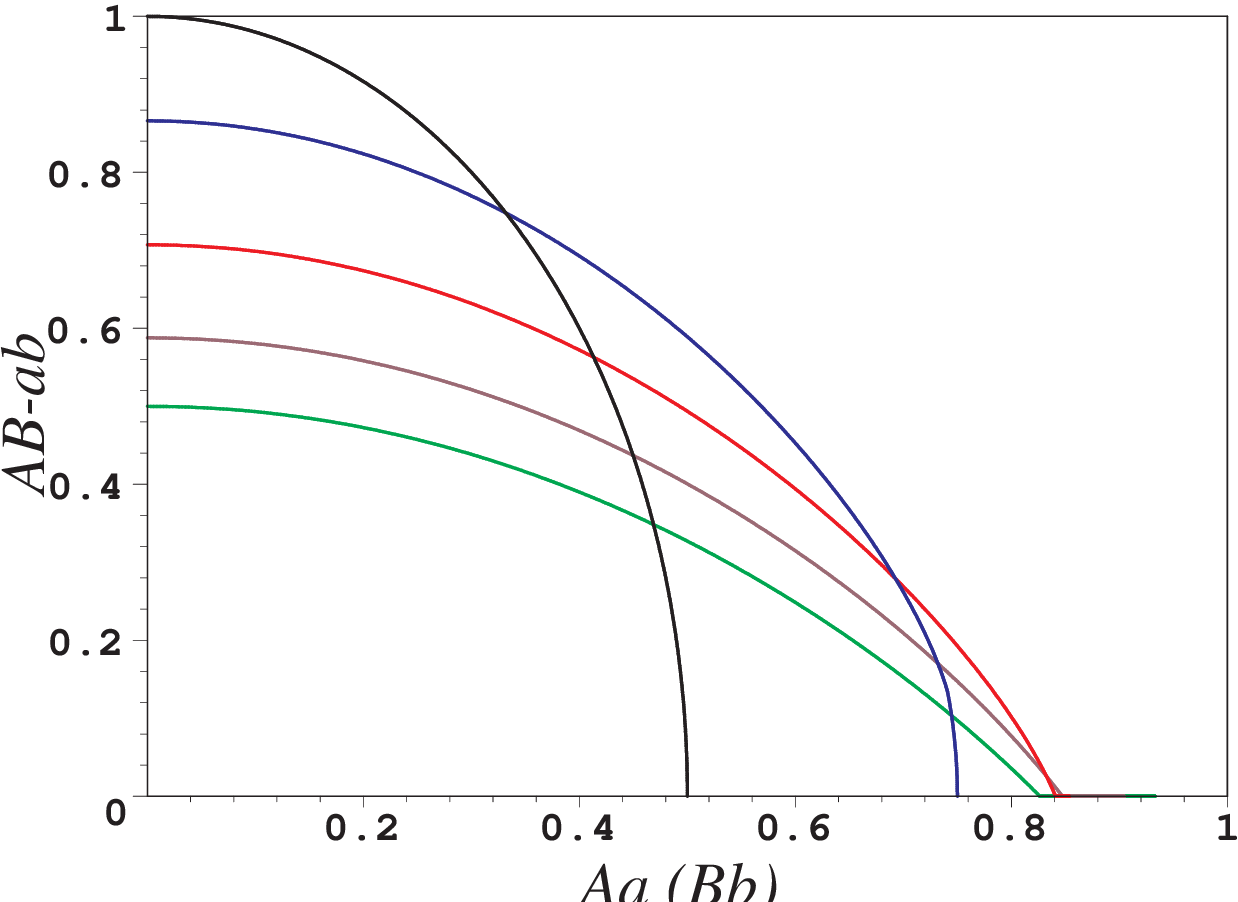}%\\
  \caption{{\footnotesize Graphic of the ellipse ${C}_{Aa(Bb)} \,\mathrm{x}\, \bigl(
  {C}_{AB}-{C}_{ab}\bigr)$
  with $\alpha=\pi/4, \pi/6, \pi/8, \pi/10 \ \mathrm{and} \ \pi/12 $ for the colors black, blue, red, brown
  and green, respectively.
  }}\label{fig11}
\end{figure}

In figure \ref{fig11}, we show that we will always have a segment of
the above ellipse, because the entanglement in
$Aa$ does not suddenly disappear. If $0<\alpha<\arctan(1/2)$ the
major semi--axis will be parallel to ${C}_{Aa(Bb)}$. When
$\alpha=\arctan(1/2)$ we have a circumference and if $\arctan(1/2)<\alpha<\pi/2$,
the major semi--axis will be parallel to ${C}_{AB}-{C}_{ab}$.
The eccentricity of (\ref{elipse_phi}) is

\begin{eqnarray}\label{excen_phi}
\bar{\epsilon}=\left\{
\begin{array}{rcl}
  \sqrt{\dfrac{5\mathcal{P}_{0}-3}{(1+\mathcal{P}_{0})}}, & \mbox{if} & 0<\alpha<\alpha_0 \\
   & & \\
  \sqrt{\dfrac{3-5\mathcal{P}_{0}}{4(1-\mathcal{P}_{0})}}, & \mbox{if} & \alpha_0<\alpha<\pi/4 \\
     & & \\
  \sqrt{\dfrac{3+5\mathcal{P}_{0}}{4(1+\mathcal{P}_{0})}}, & \mbox{if} &
  \pi/4<\alpha<\pi/2,
\end{array}
\right.
\end{eqnarray}
where $\alpha_0=\arctan(1/2)$. The focus is
\begin{eqnarray}\label{foco_phi}
\bar{f}=\left\{
\begin{array}{rcl}
  \sqrt{(5\mathcal{P}_{0}-3)(1+\mathcal{P}_{0})}/2, & \mbox{if} & 0<\alpha<\alpha_0 \\
   & & \\
  \sqrt{(3-5\mathcal{P}_{0})(1+\mathcal{P}_{0})}/2, & \mbox{if} & \alpha_0<\alpha<\pi/4 \\
     & & \\
  \sqrt{(3+5\mathcal{P}_{0})(1-\mathcal{P}_{0})}/2, & \mbox{if} &
  \pi/4<\alpha<\pi/2 \ \ \ \ .
\end{array}
\right.
\end{eqnarray}

Observe that when $\alpha=\alpha_0$ we have $\mathcal{P}_0=3/5$,
$\bar{f}=0$ and the ellipse becomes a semi circumference. Notice that
if $0<\alpha<\alpha_0$, the entanglement of $AB$ disappears before
the appearance of entanglement in $ab$. However, the entanglement of
$Aa$ is given by the JCM and does not remain zero in any finite
interval of time. As a result, there will be a time interval such
that ${C}_{AB}-{C}_{ab}$ will be zero but the entanglement between
$Aa$ will not. Thus, ${C}_{Aa}$ will admit values larger than
${C}_{AB}-{C}_{ab}$ and we have the major semi--axis parallel to
${C}_{Aa}$. 

\begin{figure}[h]

\centering
  \includegraphics[scale=0.300,angle=-90]{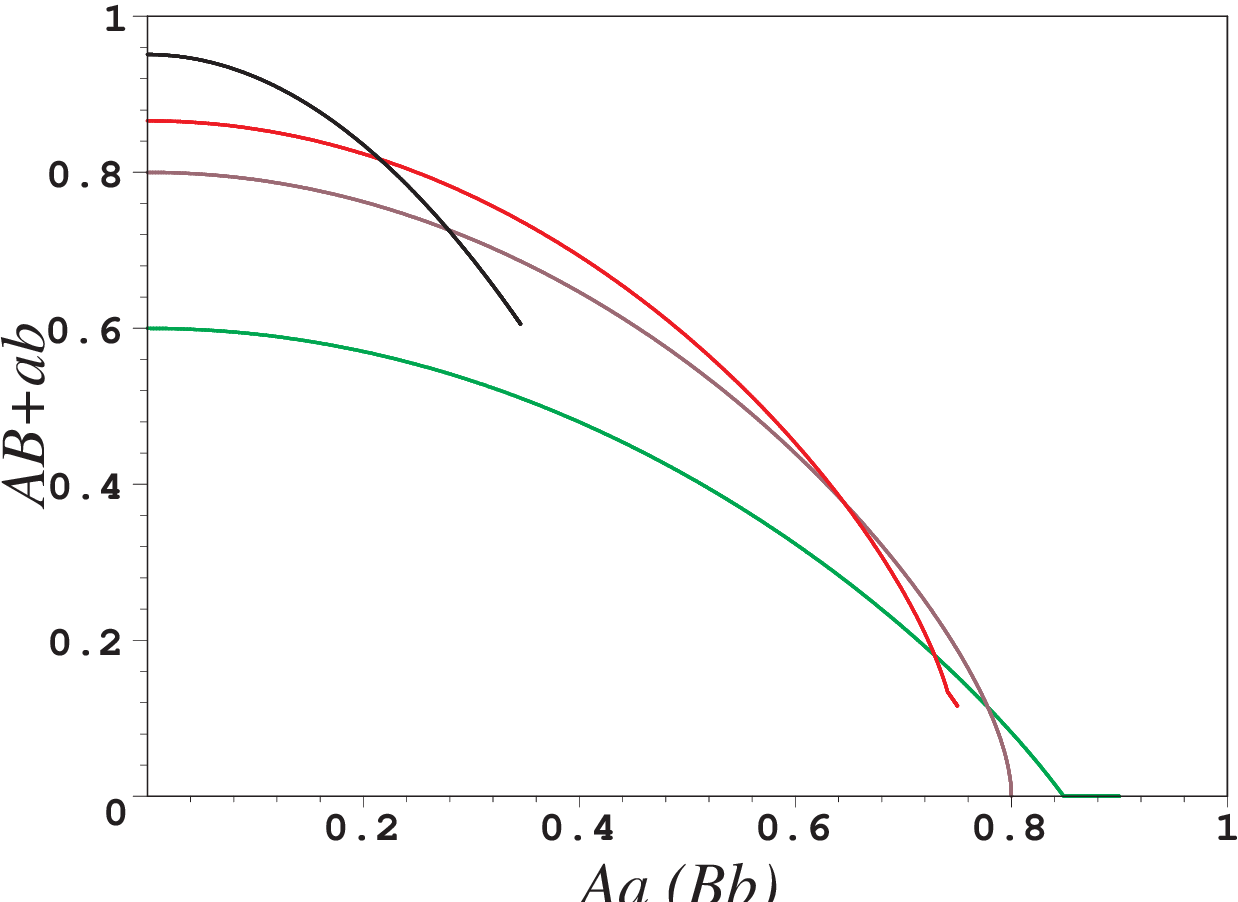}%\\
  \caption{{\footnotesize Graphic of the parabola ${C}_{Aa(Bb)} \,\mathrm{x}\, \bigl(
  {C}_{AB}+{C}_{ab}\bigr)$
  with $\alpha=3\pi/10, \pi/6, \arctan(1/2) \ \mathrm{and} \ \arctan(1/3) $ for the colors black, red, brown and green, respectively.
  }}\label{fig12}
\end{figure}

Consider now the expression used previously,
 $[C_0-({C}_{AB}+{C}_{ab})]/\cos^{2}(\alpha)=\sin^2(2gt)$.
Using equation (\ref{cc3}) or (\ref{cc6}) we get another
parabola whose equation reads
\begin{equation}\label{parabola2}
\bigl({C}_{AB}+{C}_{ab}\bigr)=C_0-\dfrac{{C}_{Aa(Bb)}^{2}}{\cos^2\alpha}\
\ \ \ .
\end{equation}
In figure \ref{fig12}, it becomes clear that the vertex $\tilde{v}$
and the focus $\tilde{f}$ are localized on the axis
$\bigl({C}_{AB}+{C}_{ab}\bigr)$ at points given by
$\tilde{v}=\{0,\sqrt{1-(\mathcal{P}_0)^2}\}$ and
$\tilde{f}_{\lessgtr}=
\{0,\sqrt{1-\mathcal{P}_0}-(1\pm\mathcal{P}_0)\}$. As before, the
sub--index is $<$ ($>$) if $0<\alpha<\pi/4$ ($\pi/4<\alpha<\pi/2$),
respectively. We always have a segment of this parabola in the
diagram $\bigl({C}_{AB}+{C}_{ab}\bigr) \,\mathrm{x}\, {C}_{Aa(Bb)}$,
because its vertex is limited between 0 and 1. We also know that
$\bigl({C}_{AB}+{C}_{ab}\bigr)$ will not be zero if
$\alpha_0<\alpha<\pi/2$ and in this interval the parabola does not
touch the axis ${C}_{Aa(Bb)}$.

\begin{figure}[h]

\centering
  \includegraphics[scale=0.300,angle=-90]{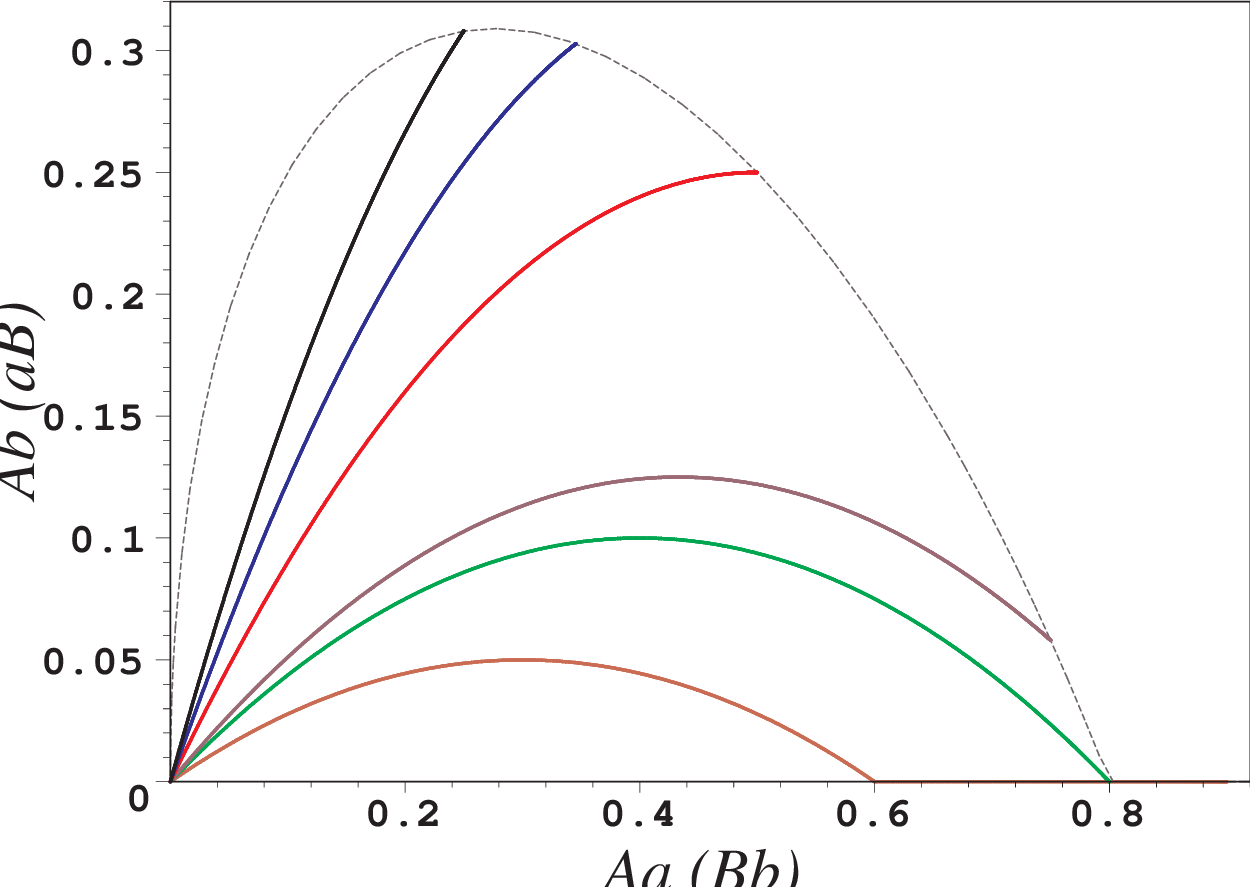}%\\
  \caption{{\footnotesize Graphic of the parabola ${C}_{Aa(Bb)} \,\mathrm{x}\, {C}_{Ab(aB)}$
  with $\alpha=\pi/3, 3\pi/10, \pi/4, \pi/6, \arctan(1/2) \ \mathrm{and} \ \arctan(1/3) $ for the colors black, red, brown, green
  and orange respectively. 
  }}\label{fig122}
\end{figure}

And last but not least, we can write $|\sin(2gt)|={C}_{Aa(Bb)}\cos^2\alpha$ 
from equations (\ref{cc3}) or (\ref{cc6}) and substitute in (\ref{cc4})
or (\ref{cc5}). With some simplifications we have
\begin{equation}\label{parabola3}
{C}_{Ab(aB)}+\dfrac{1}{2\cos^2\alpha}\biggl({C}_{Aa(Bb)}-\dfrac{C_0}{2}\biggr)^2
=\dfrac{C_{0}^{2}}{8\cos^2\alpha}.
\end{equation}
That, like the previous case, is also a parabola with
vertex $\breve{v}$ and focus $\breve{f}$
localized at points
\begin{eqnarray*}
  \breve{v}_{\lessgtr} &=& \left\{\dfrac{\sqrt{1-\mathcal{P}_{0}^{2}}}{2},\dfrac{\bigl(1\mp\mathcal{P}_0\bigr)}{2}\right\}, \\
  \breve{f}_{\lessgtr} &=&
  \left\{\dfrac{\sqrt{1-\mathcal{P}_{0}^{2}}}{2},\dfrac{\mp\mathcal{P}_0}{2}\right\}.
\end{eqnarray*}
The sub--index follows the previous notation. The parabola of
equation (\ref{parabola3}) touches twice the axis ${C}_{Aa(Bb)}$
when $0<\alpha\leq\alpha_0$. This happens because if
$0<\alpha<\alpha_0$, there is entanglement sudden death in the partition
${C}_{Ab(aB)}$. If $\alpha_0<\alpha<\pi/2$, on the other hand, there
is not sudden death and the segment of the parabola only touches the axis ${C}_{Aa(Bb)}$
at the origin, as showed in figure \ref{fig122}.

\section{The entanglement surface}\label{Sup_Emara}

\begin{figure}[h]

\centering
  \includegraphics[scale=0.300,angle=-90]{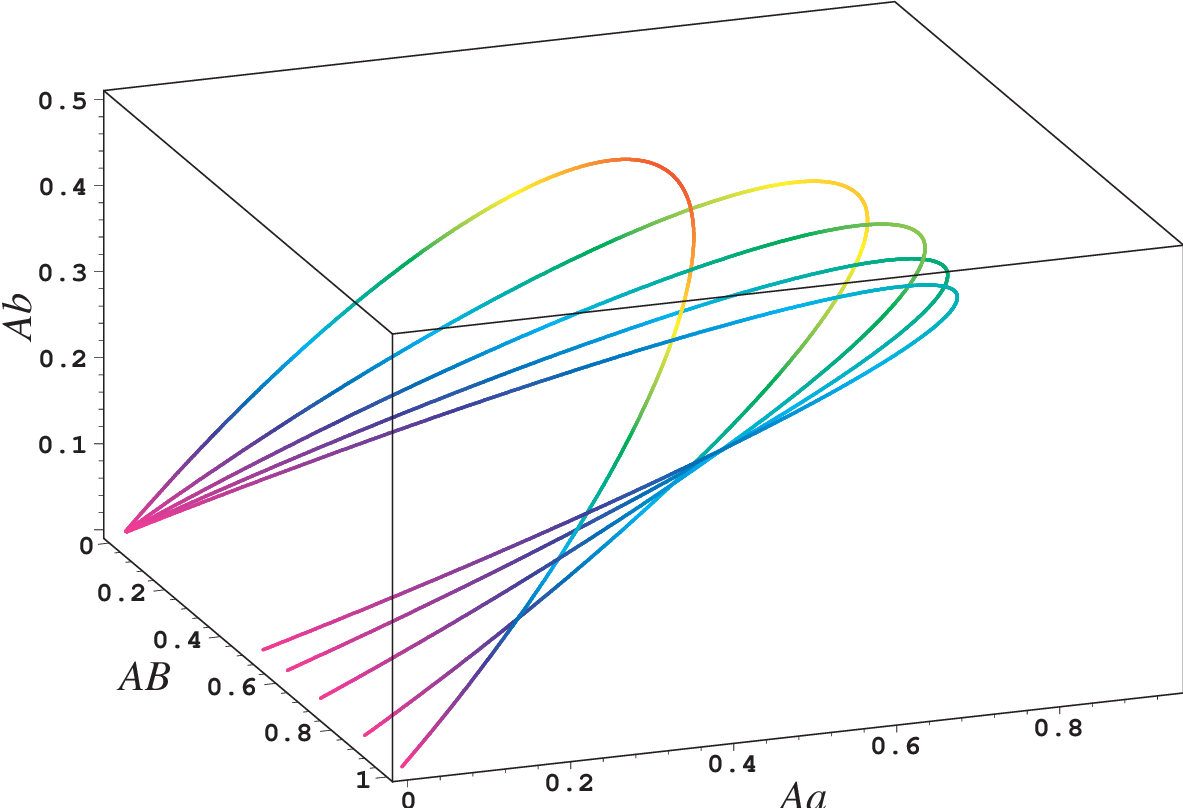}%\\
  \caption{{\footnotesize Graphic of the diagram $C_{AB} \,\mathrm{x}\, C_{Aa}\,\mathrm{x}\,
C_{Ab}$ for the atom $A$ and initial state (\ref{psi_0}). From the superior curve
to the inferior, we have, respectively, $\alpha=\pi/4, \pi/6, \pi/8, \pi/10
\ \mathrm{and} \ \pi/12$.
  }}\label{fig13}
\end{figure}

\begin{figure}[h]

\centering
  \includegraphics[scale=0.300,angle=-90]{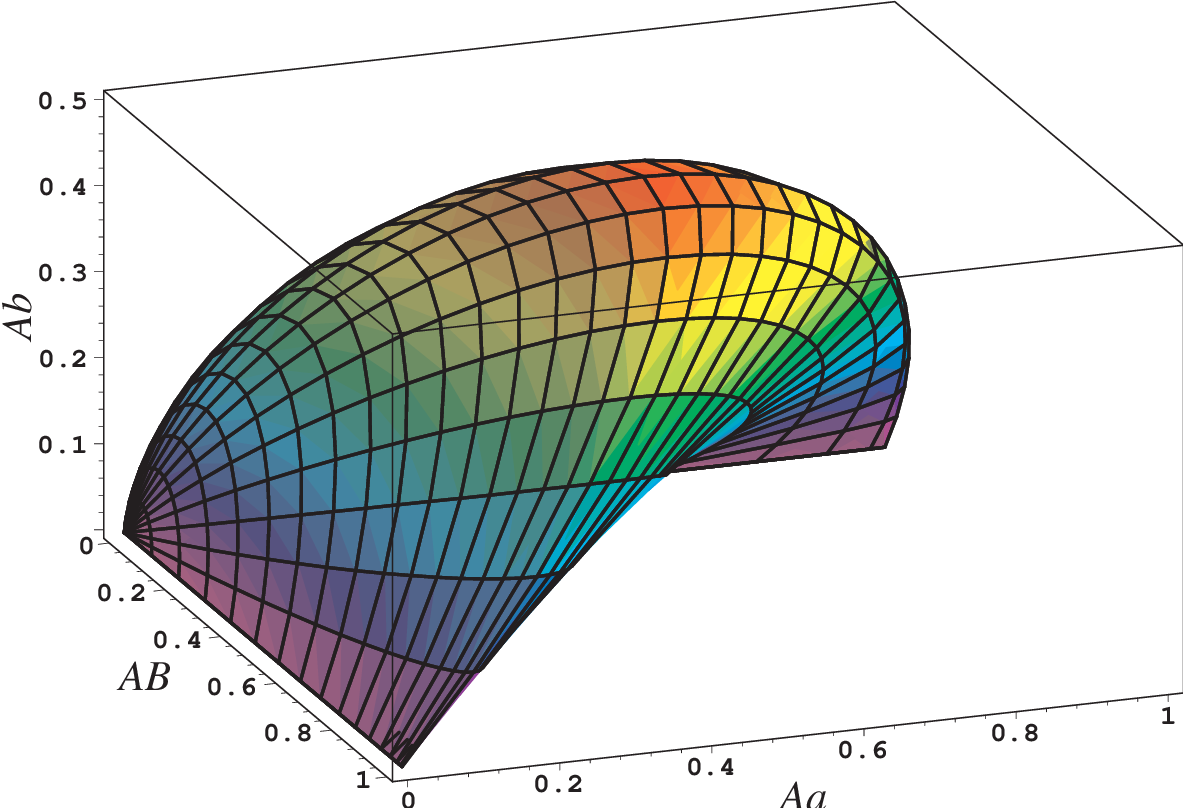}\\

  \includegraphics[scale=0.300,angle=-90]{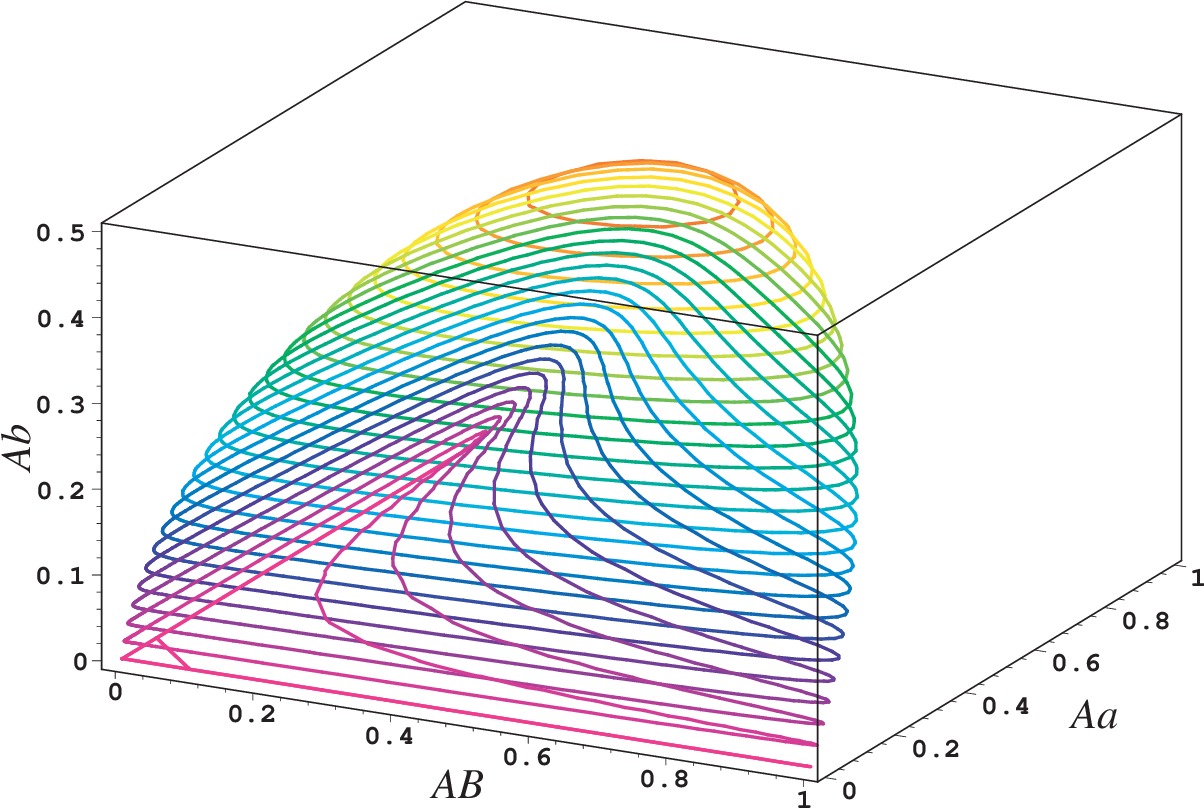}
  \caption{{\footnotesize Entanglement surface for the atom $A$ in the diagram $C_{AB} \,\mathrm{x}\, C_{Aa}\,\mathrm{x}\,
C_{Ab}$ for the initial state(\ref{psi_0}).
  }}\label{fig14}
\end{figure}

\begin{figure}[h]

\centering
  \includegraphics[scale=0.300,angle=-90]{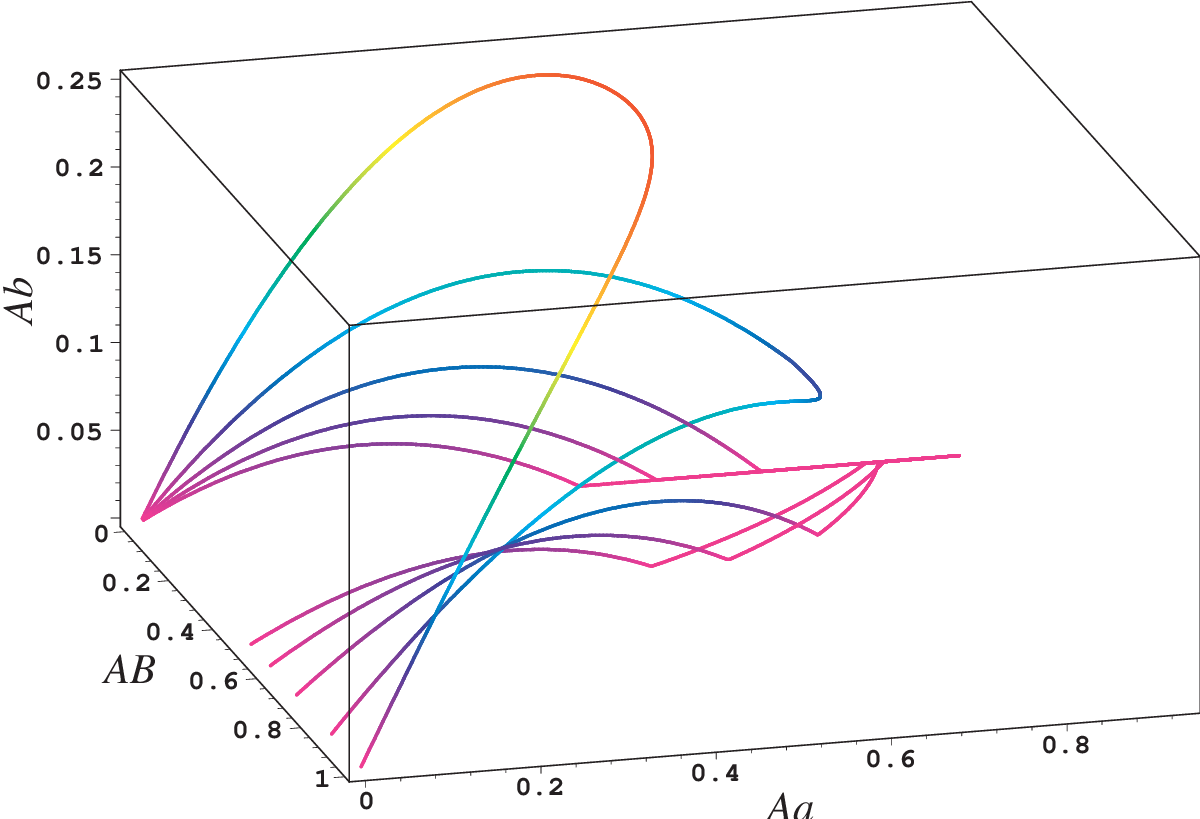}%\\
  \caption{{\footnotesize Graphic of the diagram $C_{AB} \,\mathrm{x}\, C_{Aa}\,\mathrm{x}\,
C_{Ab}$ for the initial state (\ref{phi_0}), with
$\alpha= \pi/4, \pi/6, \pi/8, \pi/10 \ \mathrm{and} \ \pi/12$.
  }}\label{fig15}
\end{figure}

In the previous section we explore the diagram $C_{ij}
\,\mathrm{x}\, C_{kl}$ for two different initial states. Because of
the unitary evolution of the physical model and the existence of
an entanglement invariant \cite{distribuicao}, it is relevant
to analyze the three dimensional diagram $C_{ij} \,\mathrm{x}\,
C_{ik}\,\mathrm{x}\, C_{il}$ for the $i$--{\it th} {\it qubit}. First we
analyze such diagram for the atom $A$. For the initial state
(\ref{psi_0}), the concurrences between the atom $A$ and any other
{\it qubit} are given by equations (\ref{c1}), (\ref{c3}) and
(\ref{c4}). If we make the parametric graphics of this concurrences
we have curves, for a determined value of $\alpha$, in a diagram
$C_{AB} \,\mathrm{x}\, C_{Aa}\,\mathrm{x}\, C_{Ab}$, as showed in
figure \ref{fig13}.
Naturally, if we look at the projections of this
curves in the planes $C_{AB} \,\mathrm{x}\, C_{Aa}$, $C_{AB}
\,\mathrm{x}\, C_{Ab}$ and $ C_{Aa}\,\mathrm{x}\, C_{Ab}$ we get
the graphics drawn in figures \ref{fig3}, \ref{fig4} and
\ref{fig5}, respectively. If we draw all the possible
curves (varying $\alpha$ from 0 to $\pi/2$) in the diagram $C_{AB}
\,\mathrm{x}\, C_{Aa}\,\mathrm{x}\, C_{Ab}$ we have a surface in that
space depicted in figure \ref{fig14}.
A point over that surface informs how much entanglement
there is in each one of the partitions $AB$, $Aa$ and $Ab$. If now
we consider the initial state (\ref{phi_0}) and draw the
parametric graphics, for a few values of
$\alpha$, in a diagram $C_{AB} \,\mathrm{x}\, C_{Aa}\,\mathrm{x}\,
C_{Ab}$, we also have curves in that diagram, as depicted in
figure \ref{fig15}. 

\begin{figure}[h]

\centering
  \includegraphics[scale=0.300,angle=-90]{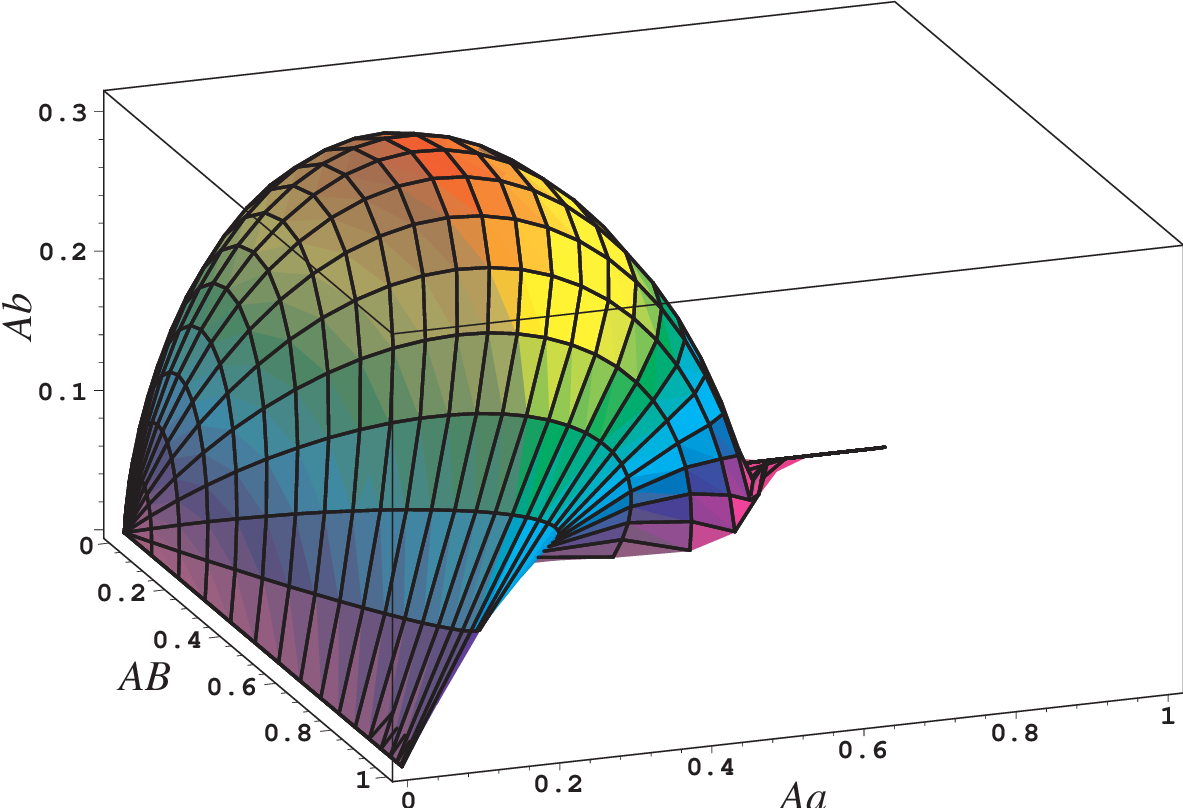}\\

  \includegraphics[scale=0.300,angle=-90]{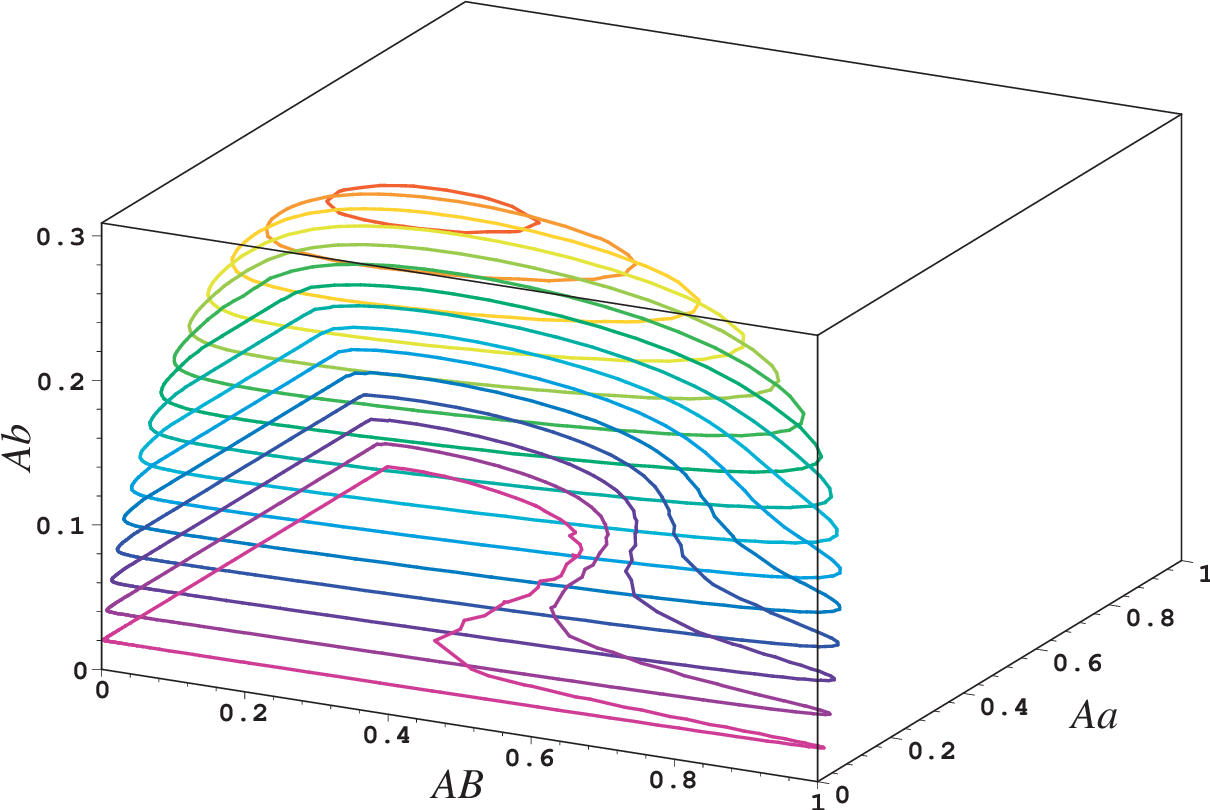}
  \caption{{\footnotesize Entanglement surface of the atom $A$ in the diagram $C_{AB} \,\mathrm{x}\, C_{Aa}\,\mathrm{x}\,
C_{Ab}$ for the initial state (\ref{phi_0}).
  }}\label{fig16}
\end{figure}

As in the previous case we can draw all possible curves
in the diagram $C_{AB} \,\mathrm{x}\, C_{Aa}\,\mathrm{x}\, C_{Ab}$
if we vary $\alpha$ from 0 to $\pi/2$ and we find an entanglement surface,
see figure \ref{fig16}.
Points over this surface also gives how much entanglement
there is in each of the subsystems $AB$, $Aa$ e $Ab$.

This same conclusions are also true for $B$, $a$ and $b$. So, in a
general way, we can say that any trajectory in the diagrams $C_{ij}
\,\mathrm{x}\, C_{ik}$, $C_{ij} \,\mathrm{x}\, C_{il}$ and $
C_{ik}\,\mathrm{x}\, C_{il}$ belongs to the surface in $C_{ij}
\,\mathrm{x}\, C_{ik}\,\mathrm{x}\, C_{il}$ and they are projections
in its respective diagrams, where $i$, $j$, $k$ and $l$ are the 4
{\it qubits} ($A$, $B$, $a$ and $b$) of the system.

\section{hyper-sphere shell of the entanglement
 dynamics}\label{hiper_esfera}

Next, we are going to use a result already obtained in
\cite{distribuicao1} and \cite{geral2}. In these references, they
observed that for the initial state (\ref{psi_0}) we have
$C_{AB}^{2}+C_{Ab}^{2}+C_{aB}^{2}+C_{ab}^{2}=C_{0}^{2}$. Without
loss of generality, we can sum in both sides the term
$C_{Aa}^{2}+C_{Bb}^{2}$ and this yields
$C_{AB}^{2}+C_{Ab}^{2}+C_{aB}^{2}+C_{ab}^{2}+C_{Aa}^{2}+C_{Bb}^{2}=C_{0}^{2}+
(\cos^4\alpha+\sin^4\alpha)\sin^2(2gt)$. This last expression can be
transformed in the inequality
$C_{AB}^{2}+C_{Ab}^{2}+C_{aB}^{2}+C_{ab}^{2}+C_{Aa}^{2}+C_{Bb}^{2}\leq
C_{0}^{2}+ (\cos^4\alpha+\sin^4\alpha)$. Now, if we use simple
trigonometric relations and the {\it predictability}, we can rewrite
this equation as
\begin{equation}\label{des_esfera1}
C_{AB}^{2}+C_{Ab}^{2}+C_{aB}^{2}+C_{ab}^{2}+C_{Aa}^{2}+C_{Bb}^{2}\leq1+\dfrac{C_{0}^{2}}{2},
\end{equation}
which is a hyper-sphere with radius $\sqrt{(2+C_{0}^{2})/2\,}$ in a
space where the axes are the concurrences between pairs of {\it qubits}.
Besides, we can generalize the above inequality to
\begin{equation}\label{casca_esfera1}
C_{0}^{2}\leq
C_{AB}^{2}+C_{Ab}^{2}+C_{aB}^{2}+C_{ab}^{2}+C_{Aa}^{2}+C_{Bb}^{2}\leq1+\dfrac{C_{0}^{2}}{2},
\end{equation}
which defines a limited region (a hyper-sphere shell) inside the
hypersphere defined by eq.(\ref{des_esfera1}). Thus, any curve in a
diagram where the axes are concurrences between pairs of {\it qubits}
and the initial state is (\ref{psi_0}) will lie either on the
surface or in the interior of the hypersphere shell
(\ref{casca_esfera1}). So, we can speculate that, in the same way
that curves in diagrams $C_{ij} \,\mathrm{x}\, C_{ik}$ are
projections of curves of $C_{ij} \,\mathrm{x}\, C_{ik}\,\mathrm{x}\,
C_{il}$, the surface defined in $C_{ij} \,\mathrm{x}\,
C_{ik}\,\mathrm{x}\, C_{il}$ is a projection of the surface of a
hypersphere that is in a space of greater dimension.

We can make the same analysis for the initial state (\ref{phi_0}).
However, in that case
\cite{distribuicao1,geral2} we have only the inequality $0\leq
C_{AB}^{2}+C_{Ab}^{2}+C_{aB}^{2}+C_{ab}^{2}\leq C_{0}^{2}$ and, as done
previously, we can sum both sides with the term
$C_{Aa}^{2}+C_{Bb}^{2}=\bigl[1+\cos(2\alpha)\bigr]^2\sin^2(2gt)/2$.
With a simple algebra we can express the result of this sum
in the inequality $0\leq C_{AB}^{2}+C_{Ab}^{2}+C_{aB}^{2}+C_{ab}^{2} +
C_{Aa}^{2}+C_{Bb}^{2} \leq C_{0}^{2}+[1+\cos(2\alpha)]^2/2$. We have
the predictability $\mathcal{P}_0$ equals to $\cos(2\alpha)$ if
$0<\alpha<\pi/4$ and $-\cos(2\alpha)$ if $\pi/4\leq\alpha\leq\pi/2$.
Using this and $C_{0}^{2}+\mathcal{P}_{0}^{2}=1$ we can rewrite the
inequality as
\begin{equation}\label{des_esfera2}
0\leq C_{AB}^{2}+C_{Ab}^{2}+C_{aB}^{2}+C_{ab}^{2} +
C_{Aa}^{2}+C_{Bb}^{2}\leq 1+ \frac{C_{0}^{2}}{2}\pm\mathcal{P}_0,
\end{equation}
where on the right hand side of the equation we will have $(1+
C_{0}^{2}/2+\mathcal{P}_0)$ when $0<\alpha<\pi/4$ and $(1+
C_{0}^{2}/2-\mathcal{P}_0)$ when $\pi/4\leq\alpha\leq\pi/2$. This
inequality must be valid during the whole evolution
and, in a space defined by the axes corresponding to the concurrences $C_{ij}$.
We have the radius of the hyper-sphere given by
\begin{equation}\label{raio}
R_{\gtrless}=\sqrt{1+ \frac{C_{0}^{2}}{2}\pm\mathcal{P}_0 \,},
\end{equation}
where we have $R_>$ ($R_<$) when $0<\alpha<\pi/4$
($\pi/4\leq\alpha\leq\pi/2$), respectively. It is noteworthy
that for $0<\alpha<\pi/4$ there is sudden death of entanglement
 in few partitions. On the other hand, for
$\pi/4\leq\alpha\leq\pi/2$ there is not sudden death for any
partition. Thus, we have $R_>$ when there is sudden death and $R_<$
otherwise. Note that for $0<\alpha<\arctan(1/2)$ there will be a
time interval
$\Delta\tau=[\arccos(\sqrt{\tan\alpha\,})-\arcsin(\sqrt{\tan\alpha\,})]/g$
where $C_{AB}=C_{ab}=C_{Ab}=C_{aB}=0$ (as observed in
\cite{nascimento}). Since the hyper-sphere is defined by the
concurrences between pairs of {\it qubits}, one would intuitively
expect, in this conditions
and during the time interval $\Delta\tau$, to obtain $R_<$ in place
of $R_>$ since only $C_{Aa}$ and $C_{Bb}$ are different from zero.
The increasing of the average radius is a consequence of the dynamical
entanglement $C_{Aa}$ and $C_{Bb}$.
When $0<\alpha<\pi/4$ the entanglement of the partitions $Aa$ and
$Bb$ will attain maximum values between 1/2 and 1. Thus, the maximum
value of $C_{Aa}^{2}+C_{Bb}^{2}$ will be between 1/2 and 2,
contributing substantially to the inequality (\ref{des_esfera2}).

\section{The effect of perturbation on conics}
\label{sec6}

In this section, we consider the effect of decoherence in order to see
how some conics of section \ref{Din_emara} are affected. We consider the two cavities
decaying freely, i. e. they both interact with a reservoir at zero temperature. This is a model
closer to experimental reality. 

The solution of the master equation for the initial state (\ref{psi_0}) gives us the density matrix
for this case. We take the partial trace over the subsystems in order to obtain the following 
concurrences

\begin{eqnarray}
C_{AB}&=&e^{-rz}|\sin2\alpha|\left[\cos\left(\frac{z}{2}\right)+r\sin\left(\frac{z}{2}\right)\right]^2, \\
C_{Aa}&=&e^{-rz}\sqrt{1+r^2}\cos^2\alpha|r(1-\cos z)+ \sin z|, \\
C_{Bb}&=&e^{-rz}\sqrt{1+r^2}\sin^2\alpha|r(1-\cos z)+ \sin z|, \\
C_{Ab}&=&\frac{1}{2} e^{-rz}\sqrt{1+r^2}|\sin 2\alpha||r(1-\cos z)+ \sin z|, \\
C_{aB}&=&\frac{1}{2} e^{-rz}\sqrt{1+r^2}|\sin 2\alpha||r(1-\cos z)+ \sin z|, \\
C_{ab}&=&e^{-rz}(1+r^2)|\sin 2\alpha|\sin^2 \left(\frac{z}{2}\right),
\end{eqnarray}
where $k$ is the decay constant, $\Omega=\sqrt{4g^2-k^2}$ is the Rabi frequency, $r=\frac{k}{\Omega}$ is
the ratio between them and $z=\Omega t$. We recover eqs. (\ref{c1})--(\ref{c6}) in the limit $k\rightarrow 0$.

\begin{figure}[h]
\centering
    \subfigure[]
    {   \includegraphics[scale=0.500,angle=0]{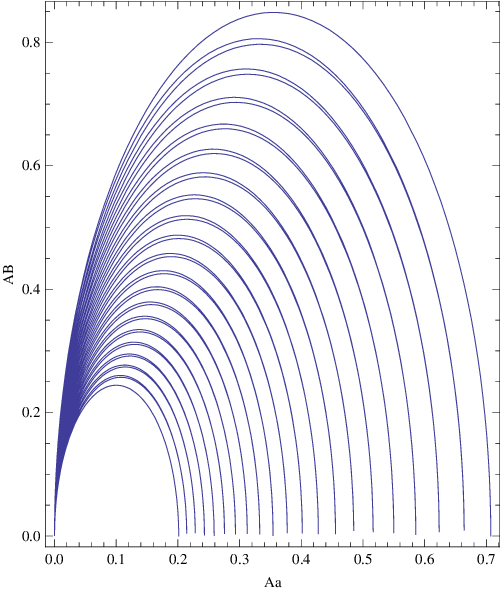}
  \label{fig19} }
    \subfigure[]
{   \includegraphics[scale=0.500,angle=0]{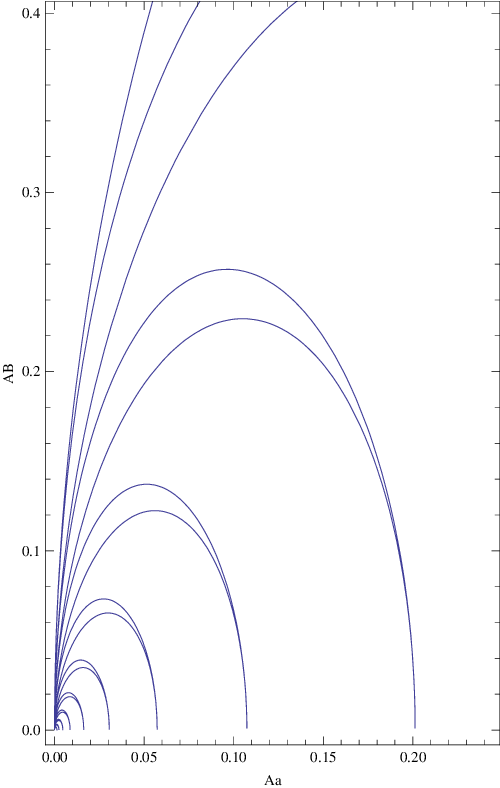}
\label{fig20}}
\caption{{\footnotesize(a) Graphic of the semi--ellipse $C_{AB} \,\mathrm{x}\, C_{Aa}$
  with $\alpha= \pi/8$, $r=0.01$ and $z \in [0,40\pi]$.(b) Graphic of the semi--ellipse $C_{AB} \,\mathrm{x}\, C_{Aa}$
  with $\alpha= \pi/8$, $r=0.1$ and $z \in [0,20\pi]$
  }}
\end{figure}

Surprisingly enough, the curves (\ref{c12}), (\ref{c13}), (\ref{elipse_aB(Ab)_Aa}) and (\ref{elipse_Ab(aB)_Bb}) are not affected 
by this type of external coupling considered. On the other hand, if we consider one of the ellipses of subsection \ref{s3.1.1}, we see
that its size decreases with time. This result shows what we would expect, i.e. the decoherence destroys the 
entanglement between the atoms and the entanglement between an atom and its cavity (see figure \ref{fig19}). Of course, this effect
also depends on how large $r$ is. Figure \ref{fig19} and \ref{fig20} shows examples where we see that the semi-axes go to zero in less
time with the increasing of the external coupling.

\begin{figure}[!h]

\centering
  \includegraphics[scale=0.600,angle=0]{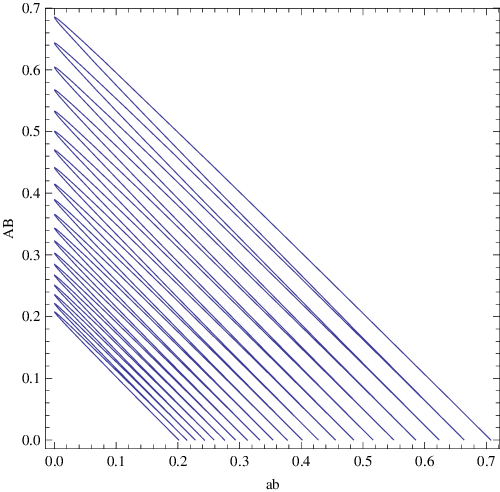}\\
  \caption{{\footnotesize Graphic of the straight line $C_{AB} \,\mathrm{x}\, C_{ab}$
  with $\alpha= \pi/8$, $r=0.01$ and $z \in [0,40\pi]$.
  }}\label{fig21}
\end{figure}

The straight line of equation (\ref{c11}) is also affected by the environment as we can see in figure \ref{fig21}. In this case, the intersection of the lines with the axes $C_{AB}$ and $C_{ab}$ shows that the initial available entanglement ($C_0$) is decreasing because the environment is monitoring the system.

Furthermore, we can notice by observing figures \ref{fig19}, \ref{fig20} and \ref{fig21} that the eccentricity of the ellipse and the angular coefficient do not change
considerably with time. Therefore, it is possible to related those with {\it predictability} as done in section \ref{Din_emara}.

\section{Conclusions}
\label{conclusion}

We presented a detailed study of the geometric character of the
entanglement dynamics of two pairs of {\it qubits} evolving according
to the DJCM. Although, this is an analytically solvable simple model,
it exhibits a very rich dynamical structure which we explored here
in order to give a geometric meaning to the entanglement dynamics.
As it became clear, its very difficult to generalize our results to
other more sophisticated models or initial conditions. However, 
we strongly believe that there is an intimate connection between
the average radius of the hypersphere and the phenomenon of sudden
death of entanglement. We hope to have provided for a tool which
might aid experimentalists given that DJCM
is within todays available technology.

\end{document}